\documentclass[a4paper, 12pt, openany, oneside]{article}
\usepackage[cp1251]{inputenc} %- для WiN
\usepackage[english, russian]{babel} %- для рус-англ. переноса
\usepackage{amsmath, latexsym, amssymb, array, graphics,
amsfonts, amsmath, bm, eucal}

\makeatletter

\newcommand{\intl}{\int\limits}
\makeatother
\usepackage{indentfirst}

\makeatother
\renewcommand{\baselinestretch}{1.2}
\usepackage[dvips]{graphicx}
\usepackage[flushleft,small]{caption2}

\begin{document}
\newcommand{\mc}[1]{\mathcal{#1}}
\newcommand{\E}{\mc{E}}
 \thispagestyle{empty}
 \renewcommand{\abstractname}{\,}
\large

\begin{flushright}\it\Large
%Dedicated to the Memory of\\our Teacher Carlo Cercignani
\end{flushright}

 \begin{center}
\bf %New method of solving of boundary problems in kinetic theory
Задача Крамерса для квантового ферми -- газа с постоянной частотой
столкновений и с зеркально -- диффузными
граничными условиями
\end{center}%\medskip

\begin{center}
  \bf
  P. V. Inanisenko\footnote{$pahanmipt@mail.ru$} and
  A. V. Latyshev\footnote{$avlatyshev@mail.ru$}
  %and  A. A. Yushkanov\footnote{$yushkanov@inbox.ru$}
\end{center}\medskip

\begin{center}
{\it Faculty of Physics and Mathematics,\\ Moscow State Regional
University, 105005,\\ Moscow, Radio str., 10--A}
\end{center}\medskip

\tableofcontents
\setcounter{secnumdepth}{4}

%\item{}\addcontentsline{toc}{chapter}{Реферат}

\begin{abstract}
The Kramers problem for quantum fermi-gases with specular -- diffuse boundary
condi\-ti\-ons of the kinetic theory is considered.
On an example of Kramers problem the new generalised method of a source  
of the decision of the boundary problems from the kinetic theory is developed.
The method allows to receive the decision with any degree of accuracy.
At the basis of a method lays the idea of representation of a
boundary condition on distribution function in the form of a source
in the kinetic equation.
By means of integrals Fourier the kinetic equation with a source
is reduced to the integral equation of Fredholm type
of the second kind.
The decision is received in the form of Neumann's series. 
\medskip
\medskip

Рассматривается классическая проблема кинетической теории для
квантового ферми--газа --
задача Крамерса с зеркально -- диффузными граничными условиями.
Задача Крамерса -- это задача о нахождении функции
распределения, массовой скорости и скорости скольжения
разреженного газа вдоль плоской твердой поверхности в случае,
когда газ движется вдоль некоторой оси, вдоль которой и
вдали от стенки задан градиент массовой скорости газа.
% \cite{LatYushk2012}.
Получено решение полупространственной
задачи Крамерса об изотермическом скольжении одноатомного газа  с постоянной
частотой столкновений молекул и с
зеркально -- диффузными граничными условиями .
Применяется новый метод решения
граничных задач кинетической теории.
Метод позволяет получить решение с произвольной степенью точности.
В основе метода лежит идея представления граничного условия на
функцию распределения в виде источника в кинетическом
уравнении. Решение получено в виде ряда Неймана.


transport theory of gases.
\end{abstract}

\begin{center}
\item{}\section{Введение. О точных решениях граничных задач
кинетической теории}
\end{center}

Задача Крамерса является одной из важнейших задач
кинетической теории газов. Эта задача имеет
большое практическое значение.
Решение этой задачи изложено в таких монографиях, как
\cite{Ferziger} и \cite{Cerc73}.

\begin{center}
  \item{}\subsection{История проблемы}
\end{center}

Более полувека тому назад К. М. Кейз в своей знаменитой работе
\cite{Case60}
заложил основы аналитического решения граничных задач теории
переноса. Идея этого метода состояла в следующем: найти общее
решение неоднородного характеристического уравнения, отвечающего
уравнению переноса, в классе обобщенных функций в виде суммы
двух обобщенных функций -- главного значения интеграла $V.P.
x^{-1}$ (valeur principal) и слагаемого, пропорционального
дельта--функции Дирака.

Первое из этих
слагаемых является частным решением неоднородного
характеристического уравнения, а второе является общим решением
соответствующего однородного уравнения, отвечающего
неоднородному характеристическому.
Коэффициентом пропорциональности в этом выражении стоит так
называемая дисперсионная функция. Нули дисперсионной функции
связаны взаимно однозначно с частными решениями исходного
уравнения переноса.

К характеристическому уравнению мы приходим
после экспоненциального (согласно Эйлеру) разделения переменных
в уравнении переноса. С помощью спектрального параметра мы
разделяем пространственную и скоростную переменные в уравнении
переноса.

Общее решение характеристического уравнения содержит в качестве
частного решения сингулярное ядро Коши $V.\, P.\, (\eta-\mu)^{-1}$,
знаменатель которого есть разность скоростной и спектральной
переменной.

Именно ядро Коши позволяет использовать всю мощь методов теории
функций комплексного переменного -- в частности, теории краевых
задач Римана --- Гильберта.

Итак, построение собственных функций характеристического
уравнения приводит к понятию дисперсионного уравнения, корни
которого находятся во взаимно однозначном соответствии с
частными (дискретными) решениями исходного уравнения переноса.

Общее решение граничных задач для уравнения переноса ищется в
виде линейной комбинации дискретных решений с произвольными
коэффициентами (эти коэффициенты называются дискретными коэффициентами)
и интеграла по спектральному параметру от собственной
функции непрерывного спектра, умноженных на неизвестную функцию
(коэффициент непрерывного спектра).
Некоторые дискретные коэффициенты задаются и считаются
известными. Дискретные коэффициенты отвечают дискретному
спектру, или, в некоторых случаях, отвечают спектру,
присоединенному к непрерывному.

Подстановка общего решения в граничные условия приводит к
интегральному сингулярному уравнению с ядром Коши. Решение этого
уравнения позволяет построить решение исходной граничной задачи
для уравнения переноса.

Действуя именно таким способом, К. Черчиньяни в 1962 г. в работе
\cite{Cerc62} построил точное решение задачи Крамерса об
изотермическом скольжении. Эта задача является важной
содержательной задачей кинетической теории.

Работы \cite{Case60, Cerc62} заложили основы аналитических
методов для получения точных решений модельных кинетических
уравнений.

Затем в работах \cite{5}--\cite{8} Черчиньяни и его соавторы
получили ряд значительных результатов для кинетической теории
газов. Эти результаты получили дальнейшее обобщение в наших
последующих работах.

Обобщение этого метода на векторный случай (системы кинетических уравнений)
наталкивается на значительные трудности (см., например,
\cite{Siewert74}). С такими трудностями столкнулись
авторы работ \cite{Siewert74, Cerc77, Aoki1}, в которых делались попытки
решить задачу о температурном скачке (задача Смолуховского).

Преодолеть эти трудности удалось в работе \cite{lat90}, в
которой впервые дано решение задачи Смолуховского.
Затем эта задача была обобщена на случай слабого испарения
\cite{ly92c} -- \cite{ly94}, не молекулярные газы \cite{ly93} и
\cite{ly98}, на безмассовые Бозе -- газы, на скачок температуры
в металле (случай вырожденной плазмы) \cite{ly03} и \cite{ly05},
и на другие проблемы \cite{ly02} и \cite{ly01}.

Затем в работах \cite{LYPMTF} и \cite{ly92b} было дано решение
задачи об умеренно сильном испарении (конденсации).
Одномерная задача о
сильном испарении была поставлена в работе \cite{Arthur} и была
сделана попытка получить ее точное решение.

Задача о температурном скачке для БГК -- уравнения с частотой
столкновений, пропорциональной модулю скорости молекул, была
решена методом Винера --- Хопфа в работе \cite{Cassell}. Затем в
более общей постановке с учетом слабого испарения (конденсации)
эта задача была решена методом Кейза в нашей работе \cite{ly96}.

Задача Крамерса в дальнейшем была обобщена на случай бинарных
газов \cite{ly91b} -- \cite{40},
была решена с использованием высших моделей
уравнения Больцмана \cite{ly97j} -- \cite{ly04s}, была обобщена на случай
зеркально -- диффузных граничных условий \cite{lyfd04} -- \cite{41}.

Нестационарные задачм для кинетических уравнений рассматривались
в наших работах \cite{ly92a} и \cite{ly98a}.

Различные проблемы теории скин -- эффекта рассматривались в
работах \cite{lyjvm99} -- \cite{44}.

В последнее десятилетие задача Крамерса была сформулирована и
аналитически решена для квантовых газов \cite{lyt03}.

Вопросам теории плазмы
посвящены наши работы \cite{ly01jv} -- \cite{53}.

В наших работах \cite{54} -- \cite{64}
были развиты приближенные методы решения граничных задач
кинетической теории с зеркально -- диффузными граничными
условиями.

В настоящей работе применяется новый эффективный метод решения
граничных задач с зеркально -- диффузными граничными условиями, развитый
в работе \cite{LatYushk2012}.

\begin{center}
  \item{}\subsection{Обобщенный метод источника}
\end{center}

В основе предлагаемого метода лежит идея включить граничное
условие на функцию распределения в виде источника в кинетическое
уравнение. Так что предлагаемый метод можно называть обобщенным 
методом источника.

Суть предлагаемого метода состоит в следующем. Сначала
формулируется в полупространcтве $x>0$ классическая задача
Крамерса об изотермическом скольжении с зеркально -- диффузными
граничными условиями. Затем функция распределения продолжается в
сопряженное полупространство $x<0$ четным образом по
пространственной и по скоростной переменным. В полупространстве
$x<0$ также формулируется задача Крамерса.

После того как получено линеаризованное кинетическое уравнение
разобьем искомую функцию (которую также будем называть
функцией распределения) на два слагаемых: чепмен ---
энскоговскую функцию распределения $h_{as}(x,\mu)$ и вторую часть функции
распределения $h_c(x,\mu)$, отвечающей непрерывному спектру:
$$
h(x,\mu)=h_{as}(x,\mu)+h_c(x,\mu),
$$
($as \equiv asymptotic, c\equiv
continuous$).

В силу того, что чепмен -- энскоговская функция распределения
есть линейная комбинация дискретных решений исходного уравнения,
функция $h_c(x,\mu)$ также является решением кинетического
уравнения. Функция $h_c(x,\mu)$ обращается в нуль вдали от
стенки. На стенке эта функция удовлетворяет зеркально --
диффузному граничному условию.

Далее мы преобразуем кинетическое уравнение для функции \\
$h_c(x,\mu)$, включив в это уравнение в виде члена типа
источника, лежащего в плоскости $x=0$, граничное условие на
стенке для функции $h_c(x,\mu)$. Подчеркнем, что функция
$h_c(x,\mu)$ удовлетворяет полученному кинетическому уравнению в
обеих сопряженных полупространствах $x<0$ и $x>0$.

Это кинетическое уравнение мы решаем во втором и четвертом
квадрантах фазовой плоскости $(x,\mu)$ как линейное
дифференциальное уравнение первого порядка, считая известным
массовую скорость газа $U_c(x)$. Из полученных решений находим граничные
значения неизвестной функции $h^{\pm}(x,\mu)$ при $x=\pm 0$,
входящие в кинетическое уравнение.

Теперь мы разлагаем в интегралы Фурье неизвестную функцию
$h_c(x,\mu)$, неизвестную массовую скорость $U_c(x)$ и дельта --
функцию Дирака. Граничные значения неизвестной функции $h_c^{\pm}(0,\mu)$
при этом выражаются одним и тем же интегралом на спектральную
плотность $E(k)$ массовой скорости.

Подстановка интегралов Фурье в кинетическое уравнение и
выражение массовой скорости приводит к характеристической
системе уравнений. Если исключить из этой системы спектральную
плотность $\Phi(k,\mu)$ функции $h_c(x,\mu)$, мы получим
интегральное уравнение Фредгольма второго рода. Ядро этого
уравнения назовем ядром Максвелла --- Неймана.

Считая градиент массовой скорости заданным, разложим неизвестную
скорость скольжения, а также спектральные плотности массовой
скорости и функции распределения в ряды по степеням коэффициента
диффузности $q$ (это ряды Неймана). На этом пути мы получаем
счетную систему зацепленных уравнений на коэффициенты рядов для
спектральных плотностей. При этом все уравнения на коэффициенты
спектральной плотности для массовой скорости имеют особенность (полюс
второго порядка в нуле). Исключая эти особенности
последовательно, мы построим все члены ряда для скорости
скольжения, а также ряды для спектральных плотностей массовой
скорости и функции распределения.

В последнем п.%араграфе
мы считаем заданным скорость скольжения, а
неизвестным мы считаем величину градиента массовой скорости.

\begin{center}
  \item{}\subsection{Изотермическое скольжение вдоль плоской поверхности}
\end{center}

Изложим физику скольжения газа вдоль плоской поверхности.
%, подробно это явление в разреженном газе вдоль плоской поверхности описано в \cite{5}.

Пусть газ занимает полупространство $x>0$ над твердой
плоской неподвижной стенкой.
Возьмем декартову систему координат с осью $x$,
перпендикулярной стенке, и с
плоскостью ($y, z$), совпадающей со стенкой, так
что начало координат лежит на стенке.

Предположим, что вдали от стенки и вдоль оси $y$
задан градиент массовой скорости газа, величина которого равна $g_v$:
$$
g_v=\left( \dfrac{d u_y(x)}{d x}\right)_{x= +\infty}.
%\eqno{(3.1)}
$$

Задание градиента массовой скорости газа вызывает
течение газа вдоль стенки.
Рассмотрим это течение в отсутствии тангенциального
градиента давления и при
постоянной температуре. В этих условиях массовая
скорость газа будет иметь
только одну тангенциальную составляющую $u_y(x)$, которая
вдали от стенки будет меняться по линейному закону. Отклонение от
линейного распределения будет
происходить вблизи стенки в слое, часто
называемом слоем Кнудсена,
толщина которого имеет порядок длины свободного
пробега $l$. Вне слоя Кнудсена течение газа описывается уравнениями
Навье --- Стокса. Явление движения газа вдоль поверхности, вызываемое
градиентом массовой скорости, заданным вдали от стенки, называется
изотермическим скольжением газа.

Для решения уравнений Навье--Стокса требуется поставить
граничные условия на
стенке. В качестве такого граничного условия принимается
экстраполированное
значение гидродинамической скорости на поверхности --
величина $u_{sl}$.

Отметим, что реальный профиль скорости в слое Кнудсена
отличен от гидродинамического. Для получения величины $u_{sl}$
требуется решить уравнение Больцмана в слое Кнудсена. При малых
градиентах скорости имеем:
$$
u_{sl}=K_{v}l G_v, \qquad
G_v=\left( \dfrac{du_y(x)}{dx}\right)_{x=+ \infty}.
%\eqno{(3.2)}
$$

Задача нахождения скорости изотермического скольжения $u_{sl}$
называется
задачей Крамерса (см., например, \cite{Ferziger}.
Определение величины $u_{sl}$
позволяет, как увидим ниже, полностью построить функцию
распределения газовых молекул в данной задаче, найти профиль
распределения в полупространстве массовой скорости газа, а
также найти значение массовой скорости газа на границе
полупространства.

Настоящая работа посвящена изучению влияния квантовых эффектов на
кинетические явления в разреженных ферми--газах. Рассмотрение ведется на
примере классической задачи об изотермическом скольжении газа
(задача Крамерса) вдоль плоской поверхности [1]--[3].
Рассматриваются как диффузные граничные условия, так и
зеркально -- диффузные граничные условия Максвелла.

Граничные условия, описывающие взаимодействие молекул газа с
поверхностью конденсированной фазы, приблекают внимание
исследователей в течение длительного времени. Эта проблема по-прежнему остается открытой, в
частности, для реальных поверхностей. В конкретных задачах
используются главным образом модельные граничные условия. Одно
из таких условий --- это зеркально -- диффузные граничные
условия Максвелла. Все параметры отраженных молекул в задачах
скольжения определяются при этом одной величиной ---
коэффициентом зеркальности, который часто отождествляют с
коэффициентом аккомодации тангенциального импульса молекул.

Условия Максвелла неплохо зарекомендовали себя при решении
конкретных граничных задач.

При наличии вдали от поверхности градиента тангенциальной к
поверхности компоненты скорости газа возникает скольжение газа
вдоль поверхности. Такое скольжение называется изотермическим
[1]--[3]. Задача Крамерса (см. [1]--[8]) состоит в нахождении
скорости изотермического скольжения газа.

Пусть газ занимает полупространство $x>0$
над плоской твердой стенкой и движется вдоль оси $y$ со средней (массовой)
скоростью $u_y(x)$ . Вдали от поверхности на расстоянии много
большем средней длины свободного пробега частиц газа
имеется градиент массовой (средней) скорости газа
$$
g_v=\Big(\dfrac{du_y(x)}{dx}\Big)_{x\to +\infty},
$$
т.е. профиль массовой скорости вдали от стенки можно представить в виде
$$
u_y(x)=u_{sl}+g_vx,\;\qquad x\to +\infty.
$$

Наличие градиента массовой скорости вызывает скольжение газа
вдоль поверхности, называемое изотермическим.
Величина $u_{sl}$ называется
скоростью изотермического скольжения ($sl\equiv sliding\equiv$
скольжение).

При малых градиентах $g_v$ скорость изотермического скольжения
пропорциональна величине градиента:
$$
u_{sl}=C_mlg_v.
\eqno{(1.1)}
$$
Здесь $C_m$ -- коэффициент изотермического скольжения,
$l$ -- средняя длина свободного пробега частиц.

Величина $C_m$ определяется кинетическими процессами вблизи
поверхности. Для ее определения необходимо решить кинетическое уравнение
в так называемом слое Кнудсена, т.е. в слое газа, примыкающего к поверхности,
толщиной порядка длины свободного пробега $l$.

В качестве кинетического уравнения рассмотрим обобщение на квантовый случай
БКВ--уравнения (Больцман, Крук, Веландер). Отметим, что в отечественной
литературе это уравнение называют также БГК--уравнением (Бхатнагар, Гросс,
Крук). Это уравнение имеет следующий вид [1--3]

$$
\dfrac{\partial f}{\partial t}+({\bf v}\nabla f)=\nu (f_{eq}-f).
\eqno{(1.2)}
$$

Здесь $f$ -- функция распределения молекул по скоростям, {\bf v} --
скорость молекул, $\nu$ -- эффективная частота столкновений молекул,
 $f_{eq}$ -- локально равновесная функция распределения,
 $$
f_{eq}=n\left(\dfrac{m}{2\pi kT}\right)^{3/2}\exp \left[-
\dfrac{m({\bf v}-{\bf u})^2}{2kT}\right].
$$

Величины $n,\;T,\;{\bf u}$ зависят, вообще говоря,
от координаты ${\bf r}$ и определяются как
$$
n=\int f d^3v,
\eqno{(1.3)}
$$
$$
{\bf u}=\dfrac{1}{n_{eq}}\int {\bf v}f d^3v,
\eqno{(1.4)}
$$
$$
T=\dfrac{2}{3kn}\int \dfrac{m}{2}({\bf v}-{\bf u})^2 f d^3v
\eqno{(1.5)}
$$

Числовая плотность (концентрация) $n$ квантового газа и его температура
$T$ в задаче Крамерса считаются постоянными.

\begin{center}
  \item{}\section{Кинетическое уравнение для квантовых ферми--газов
с постоянной частотой столкновений}
\end{center}

\begin{center}
  \item{}\subsection{Вывод уравнения}
\end{center}

Рассмотрим обобщение кинетического уравнения (1.2) на случай квантового
ферми--газа. Функцию $f_{eq}$ в (1.2) теперь будем понимать как
локально -- равновесную функцию Ферми
$$
f_{eq}=\dfrac{1}{\exp\left(\dfrac{\E_*-\mu}{kT}\right)+1},
\quad
\E_*=\dfrac{m}{2}({\bf v}-{\bf u})^2.
$$
Здесь $\mu$ -- химический потенциал молекул [8].

Вместо соотношений (1.3)--(1.5) теперь имеем следующие соотношения,
вытекающие из законов сохранения числа частиц, импульса и энергии:
$$
\int f_{eq}R_i d\Omega=\int f R_i d\Omega.
\eqno{(2.1)}
$$
Здесь
$$
d\Omega=\dfrac{(2s+1)m^3}{(2\pi \hbar)^3}d^3v,
$$
$s$ -- спин ферми--частицы,
$
R_1=1,\;R_2=v_x,\;R_3=v_y,\;R_4=v_z,\;R_5=\E_*.
$

Рассмотрим теперь применение кинетического уравнения (1.2) к задаче о
вычислении скорости изотермического скольжения квантового ферми--газа. При
этом ограничимся рассмотрением малых градиентов { $g_v$}, что позволяет
линеаризовать задачу. В этом случае температура
и концентрация газа постоянны. Из соотношений (2.1) следует, что
величина $\mathbf{u}$ совпадает с массовой скоростью газа (1.4).
Кроме того, течение газа предположим стационарным.

Линеаризуем задачу относительно равновесной
функции распределения Ферми --- Дирака (фермиана)
$f_F$

$$
f_F=\dfrac{1}{\exp\left(\dfrac{\E-\mu}{kT}\right)+1}, \qquad
\E=\dfrac{mv^2}{2}.
$$

Начнем с линеаризации локально равновесной функции распределения
$f_{eq}$. Ее линеаризуем относительно фермиана $f_F$ по массовой
скорости $\mathbf{u}$:
$$
f_{eq}=f_{eq}\Big|_{\mathbf{u}=0}+\dfrac{\partial f_{eq}}{\partial
\mathbf{u}}\Big|_{\mathbf{u}=0}\cdot\mathbf{u},
$$
что приводит к выражению
$$
f_{eq}=f_F(v)+g_F(v)\dfrac{mv_y}{kT}u_y,
\eqno{(2.2)}
$$
в котором $f_F(v)$ -- абсюлютный фермиан (см. рис. 1),
$$
f_F(v)=\dfrac{1}{1+\exp\Big(\dfrac{mv^2}{2kT}-\dfrac{\mu}{kT}\Big)}
$$
и
$$
g_F(v)=\dfrac{\exp\Big(\dfrac{mv^2}{2kT}-\dfrac{\mu}{kT}\Big)}
{\Big[1+\exp\Big(\dfrac{mv^2}{2kT}-\dfrac{\mu}{kT}\Big)\Big]^2}.
$$

Функция $g_F(v)$ называется функцией Эйнштейна (см. рис. 2).

\begin{figure}[h]
\begin{center}
\includegraphics[width=14cm,height=9cm]{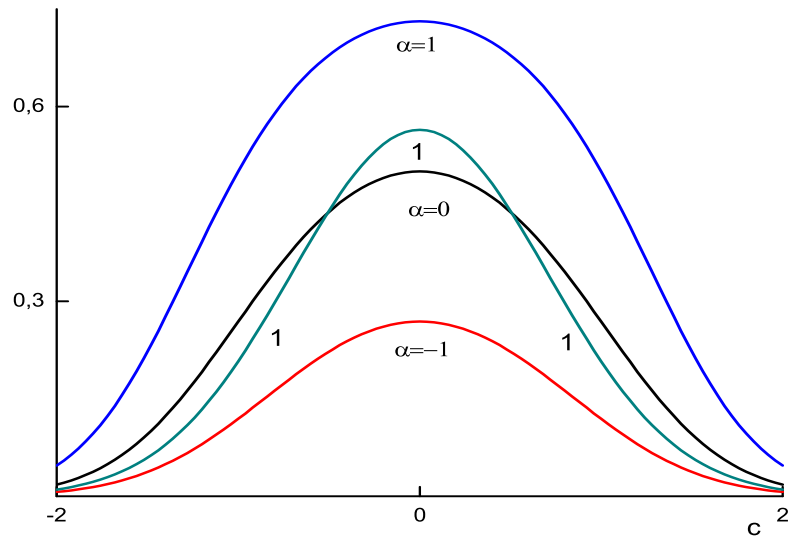}
\end{center}
\begin{center}
\caption{Поведение абсолютного фермиана (при значениях безразмерного
химпотенциала $\alpha=0,1,-1$) и абсолютного максвеллиана
$f_M(c)=\exp(-c^2)/\sqrt{\pi}$ (кривая 1)}.
\end{center}
\end{figure}

\begin{figure}[h]
\begin{center}
\includegraphics[width=14cm,height=10cm]{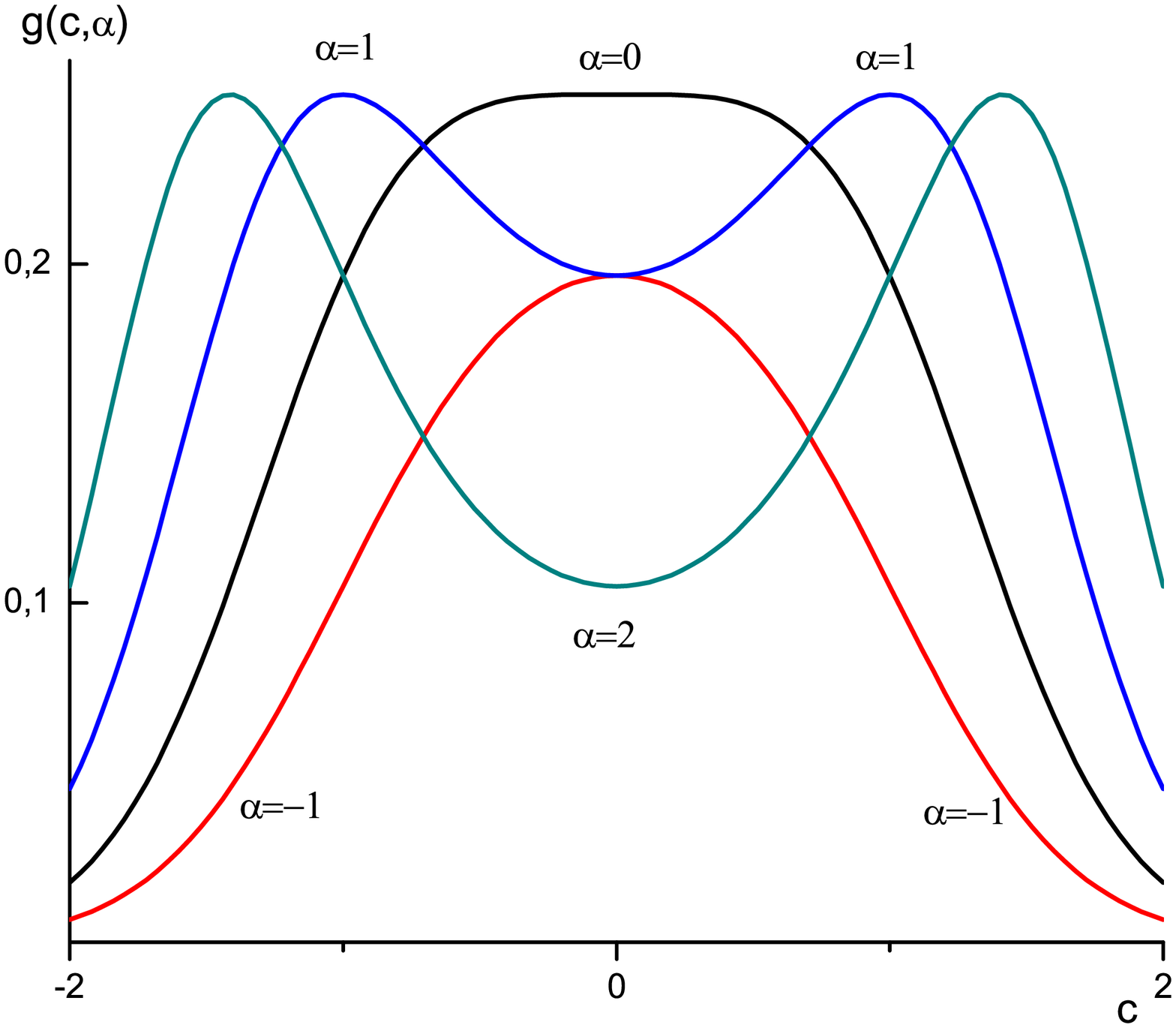}
\end{center}
\begin{center}
\caption{Поведение функции $g_F(c,\alpha)=\frac{\partial f_F(c,\alpha)}
{\partial \alpha}$ (при значениях безразмерного
химпотенциала $\alpha=0,1,-1,2$)}.
\end{center}
\end{figure}

Введем безразмерную скорость $\mathbf{C}=\sqrt{\beta}\mathbf{v},
\beta=\dfrac{m}{2kT}$, и безразмерный (приведенный) химический потенциал
$\alpha=\dfrac{\mu}{kT}$. В этих переменных выражение (2.2) записывается как
$$
f_{eq}=f_F(C)+2g_F(C)C_yU_y(x),
\eqno{(2.3)}
$$
при этом в (2.3) $U_y(x)=\sqrt{\beta}u_y(x)$ -- безразмерная
массовая скорость,
$$
f_F(C)=\dfrac{1}{1+\exp(C^2-\alpha)}, \qquad
g_F(C)=\dfrac{\exp(C^2-\alpha)}{\big[1+\exp(C^2-\alpha)\big]^2},
$$
(графики этих функций см. на рис. 1 и 2).

Согласно (2.3) функцию распределения будем искать в виде
$$
f=f(x,\mathbf{C})=f_F(C)+g_F(C)C_yh(x,C_x).
\eqno{(2.4)}
$$

Подставляя (2.3) и (2.4) в уравнение (1.2), приходим к уравнению
$$
C_x\dfrac{\partial h}{\partial x}=\nu
\sqrt{\beta}(2U_y(x)-h(x,C_x)).
$$

Далее удобно ввести безразмерную координату
$x_1=x\nu\sqrt{\beta}$.
Получим следующее уравнение
$$
C_x\dfrac{\partial h}{\partial x_1}=2U_y(x_1)-h_1(x_1,C_x).
\eqno{(2.5)}
$$

Размерный градиент $g_v=\Big(\dfrac{du_y(x)}{dx}\Big)_{x\to +\infty}$
при переходе к безразмерной координате $x_1$ преобразуется
следующим образом:
$$
G_v=\Big(\dfrac{dU_y(x_1)}{dx_1}\Big)_{x_1\to +\infty}=\sqrt{\beta}
\Big(\dfrac{du_y}{dx}\Big)_{x\to+\infty}\cdot\dfrac{dx}{dx_1},
$$
откуда
$$
G_v=\dfrac{g_v}{\nu}.
%\eqno{(2.6)}
$$

Массовая скорость газа может быть найдена из закона сохранения
импульса (2.1), который в безразмерных параметрах имеет вид
$$
\int C_y(f_{eq}-f)d\Omega=0,
$$
или,
$$
\int C_y^2g_F(C)\Big[2U_y(x_1)-h(x_1,C_x)\Big]d^3C=0,
$$
откуда получаем
$$
U_y(x_1)=\dfrac{\displaystyle\int C_y^2g_F(C)h(x_1,C_x)d^3C}
{\displaystyle 2\int C_y^2g_F(C)d^3C}.
\eqno{(2.6)}
$$

Заметим, что после линеаризации массовой скорости (1.4), т.е.
после подстановки в (1.4) выражения (2.4), приходим к следующему
выражению
$$
U_y(x_1)=\dfrac{\displaystyle\int C_y^2g_F(C)h(x_1,C_x)d^3C}
{\displaystyle \int f_F(C)d^3C}.
\eqno{(2.7)}
$$

Знаменатели в (2.6) и (2.7) имеют различные выражения. Покажем,
что эти выражения совпадают.

Перейдем к сферическим координатам
$$
C_x=C\cos \theta,\qquad C_y=C\sin \theta\sin\varphi,\qquad
C_z=C\sin \theta \sin \varphi,\qquad
$$
$$
d^3C=C^2\sin\theta d\varphi dC.
$$

Теперь получаем, что знаменатель из (2.7) равен
$$
\int f_F(C)d^3C=4\pi \int \dfrac{C^2dC}{1+\exp(C^2-\alpha)}=
4\pi f_2^F(\alpha),
$$
где $f_2^F(\alpha)$ -- второй (второго порядка) полупространственный момент
абсолютного фермиана
$$
f_2^F(\alpha)=\int\limits_{0}^{\infty}f_F(C)C^2dC.
$$

После интегрирования по частям получаем
$
f_2^F(\alpha)=\dfrac{1}{2}l_0^F(\alpha),
$
где
$$
l_0^F(\alpha)=\int\limits_{0}^{\infty}\ln(1+\exp(\alpha-C^2))dC=
\dfrac{1}{2}\int\limits_{-\infty}^{\infty}\ln(1+\exp(\alpha-C^2))dC.
$$

Знаменатель из (2.6) равен
$$
2\int C_y^2 g_F(C)d^3C=\dfrac{8\pi}{3}g_4^F(\alpha),
$$
где
$$
g_4^F(\alpha)=\dfrac{3}{2}f_2^F(\alpha).
$$
Следовательно, знаменатель из (2.6) равен:
$$
2\int C_y^2
g_F(C)d^3C=\dfrac{8\pi}{3}\cdot\dfrac{3}{2}f_2^F(\alpha),
$$
что и означает совпадение формул (2.6) и (2.7).

В числителе (2.6) удобнее перейти к цилиндрическим координатам, полагая
$C^2=C_x^2+C_{\bot}^2$, $C_y=C_{\bot}\sin\varphi$,
$d^3C=C_{\bot}dC_{\bot}dC_xd\varphi$.
Далее получаем:
$$
\int C_y^2g_F(C)h(x_1,C_x)d^3C=\int\limits_{-\infty}^{\infty}
h(x_1,C_x)dC_x\int\limits_{0}^{\infty}C_{\bot}^3g_F(C)dC_{\bot}
\int\limits_{0}^{2\pi}\cos^2\varphi d\varphi=
$$
$$
=\pi\int\limits_{-\infty}^{\infty}h(x_1,C_x)dC_x\int\limits_{0}^{\infty}
\dfrac{\exp(C_x^2+C_{\bot}^2-\alpha)C_{\bot}^3dC_{\bot}}
{[1+\exp(C_x^2+C_{\bot}^2-\alpha)]^2}.
$$

Вычисляя внутренний интеграл по частям, имеем:
$$
\int
C_y^2g_F(C)h(x_1,C_x)d^3C=\dfrac{\pi}{2}\int\limits_{-\infty}^{\infty}
\ln(1+\exp(\alpha-C_x^2))h(x_1,C_x)dC_x.
$$

Следовательно, массовая скорость вычисляется по формуле
$$
U_y(x_1)=\dfrac{1}{4l_0^F(\alpha)}\int\limits_{-\infty}^{\infty}
\ln(1+\exp(\alpha-C_x^2))h(x_1,C_x)dC_x.
\eqno{(2.8)}
$$

Введем функцию
$$
K_F(\mu,\alpha)=\dfrac{\ln(1+\exp(\alpha-\mu^2))}{2l_0^F(\alpha)}=
\dfrac{\ln(1+\exp(\alpha-\mu^2))}{\int\limits_{-\infty}^{\infty}
\ln(1+\exp(\alpha-\tau^2))d\tau},
\eqno{(2.9)}
$$
где $\mu=C_x$.

Эта функция обладает свойством
$$
\int\limits_{-\infty}^{\infty}K_F(\mu,\alpha)d\mu \equiv 1, \qquad
\forall \alpha\in(-\infty,+\infty).
%\eqno{(2.10)}
$$

Семейство функций
$K_F(\mu,\alpha)=\dfrac{\ln(1+e^{\alpha-\mu^2})}{2l_0^F(\alpha)}$
называется ядром кинетического уравнения (см. рис. 3.3).

%\clearpage
\begin{figure}[h]
\begin{center}
\includegraphics[width=16cm,height=10cm]{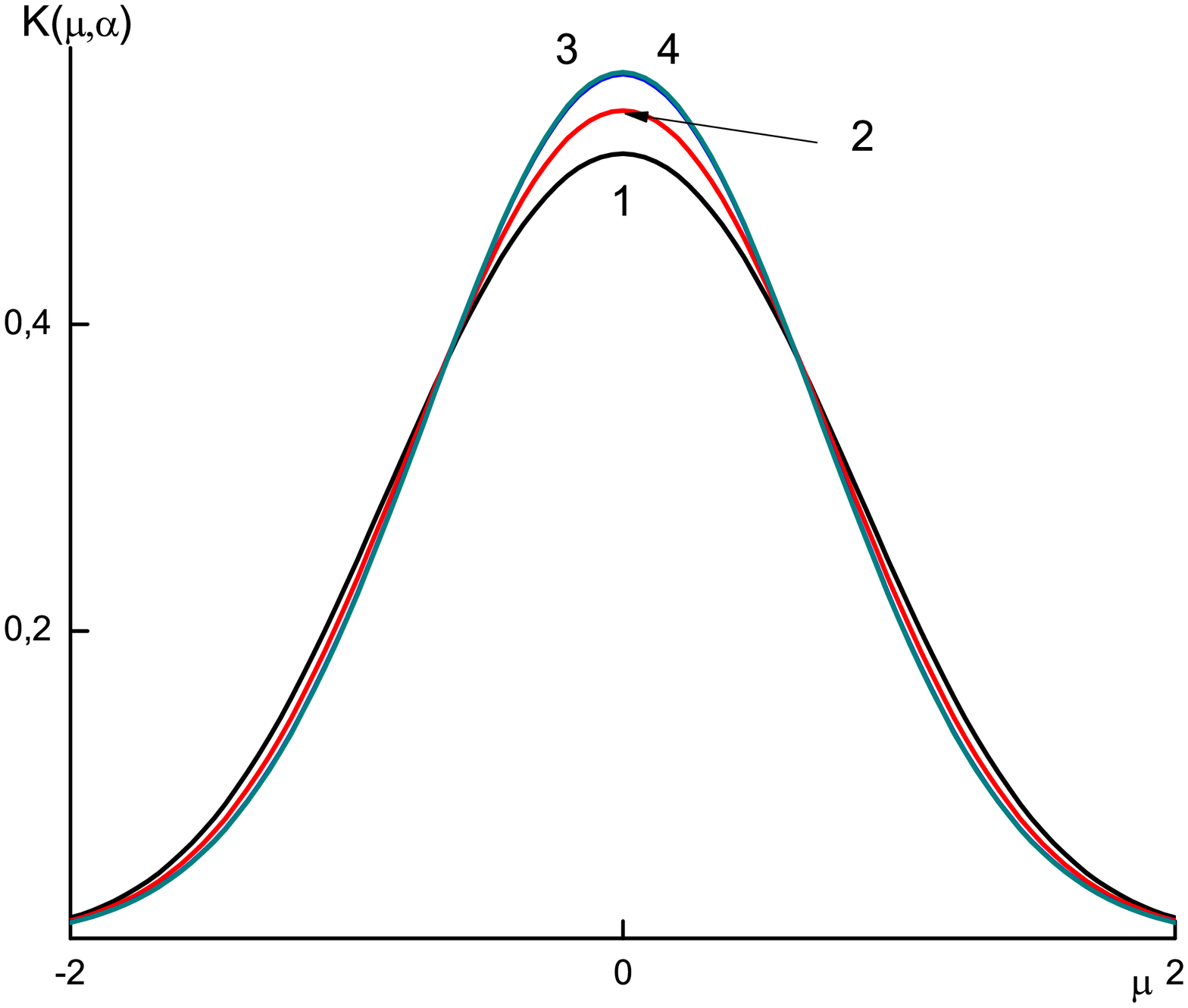}
\end{center}
\begin{center}
\caption{Поведение ядер кинетического уравнения $K_F(\mu,\alpha)$
для ферми--газа при различных значениях безразмерного
химпотенциала. Кривая 1 отвечает значению $\alpha=0$, кривая 2 --
 значению $\alpha=-1$, кривая 3, отвечающая значению $\alpha=-4$
фактически совпадает с абсолютным максвеллианом $f_M(\mu)=\exp(-\mu^2)
/\sqrt{\pi}$ (кривая 4).}
\end{center}
\end{figure}
%\clearpage

Массовая скорость согласно (2.8) и (2.9) равна
$$
U_y(x_1)=\dfrac{1}{2}\int\limits_{-\infty}^{\infty}K_F(\mu',\alpha)
h(x_1,\mu')d\mu'.
\eqno{(2.10)}
$$
Таким образом, согласно (2.10) уравнение (2.5) представим в
стандартном для теории переноса виде:
$$
\mu\dfrac{\partial h}{\partial x_1}+h(x_1,\mu)=
\intl_{-\infty}^\infty K_F(\mu';\alpha)h(x_1,\mu')\,d\mu',
\eqno{(2.11)}
$$
или, в явном виде
$$
\mu\dfrac{\partial h}{\partial x_1}+h(x_1,\mu)=
\dfrac{1}{2l_0^F(\alpha)}\intl_{-\infty}^\infty
\ln(1+\exp(\alpha-{\mu'}^2))h(x_1,\mu')\,d\mu'.
\eqno{(2.12)}
$$

\begin{center}
\item{} \subsection{Предельные случаи уравнения}
\end{center}

Рассмотрим два предельных случая уравнения (2.12): при $\alpha\to
-\infty$ и при $\alpha\to +\infty$. В первом случае
$$
\lim_{\alpha \to -\infty}K_F(\mu,\alpha)=\lim_{\alpha \to -\infty}
\dfrac{\ln(1+\exp(\alpha-\mu^2))}
{\intl_{-\infty}^\infty \ln(1+\exp(\alpha-u^2))\,du}=
$$
 $$
=\dfrac{\exp(-\mu^2)}{\intl_{-\infty}^\infty \exp(-u^2)\,du}=
\dfrac{\exp(-\mu^2)}{\sqrt{\pi}}
$$
и мы получаем уравнение
 $$
\mu\dfrac{\partial h}{\partial x}+h(x_1,\mu)=
\dfrac{1}{\sqrt{\pi}}\intl_{-\infty}^\infty\exp(-{\mu'}^2)h(x_1,\mu')\,d\mu',
$$
которое является БКВ--уравнением для одноатомных газов [1--3] с постоянной
частотой столкновений молекул.

Рассмотрим второй случай. При больших $\alpha$ ($\alpha\gg
1$) имеем:
$$
\ln(1+(\alpha-u^2))=\left\{\begin{array} {l}\alpha-u^2,\;|u|<\sqrt{\alpha},\\
0,\;|u|>\sqrt{\alpha},\end{array}\right.
$$

Следовательно, уравнение (2.12) при $\alpha\gg 1$ имеет вид:
 $$
\mu\dfrac{\partial h}{\partial x}+h(x_1,\mu)=
\intl_{-\sqrt{\alpha}}^{\sqrt{\alpha}}
\dfrac{(\alpha-{\mu'}^2)h(x_1,{\mu'}^2)}
{\intl_{-\sqrt{\alpha}}^{\sqrt{\alpha}}(\alpha-y^2)\,dy}\,d\mu'.
$$
После замены переменных
$$\mu\to \sqrt{\alpha}\mu,\qquad \mu'\to \sqrt{\alpha}\mu',\qquad
u\to \sqrt{\alpha}u,\qquad u\to \sqrt{\alpha}u
$$
получаем уравнение
 $$
\sqrt{\alpha}\mu\dfrac{\partial h}{\partial x_1}+h(x,\sqrt{\alpha}\mu)=
\dfrac{3}{4}\intl_{-1}^{1}(1-{\mu'}^2)h(x_1,\sqrt{\alpha}\mu')\,d\mu'.
$$
Изменив масштаб $x'=\sqrt{\alpha}x_1$ и обозначив
 $h_1(x',\mu)=h(\sqrt{\alpha}x_1,\sqrt{\alpha}\mu)$,
приходим к уравнению
 $$
\mu\dfrac{\partial h_1}{\partial x'}+h_1(x',\mu)=
\dfrac{3}{4}\intl_{-1}^1(1-{\mu'}^2)h_1(x',{\mu'})\,d\mu'.
$$
Это уравнение является (см. [8]) БКВ--уравнением для одноатомных газов с
частотой столкновений, пропорциональной молекулярной скорости.

\begin{center}
  \item{}\subsection{Постановка задачи Крамерса}
\end{center}

Задание градиента массовой скорости (2.2) означает, что вдали от
стенки распределение массовой скорости в полупространстве имеет
линейный рост
$$
u_y(x)=u_{sl}(\alpha)+g_vx, \qquad x\to +\infty,
$$
где $u_{sl}(\alpha)$ -- неизвестная скорость скольжения.

Умножая это равенство на $\sqrt{\beta}$ и учитывая связь
размерного и безразмерного градиентов $g_v=\nu G_v$, для
безразмерной массовой скорости получаем
$$
U_y(x_1)=U_{sl}(\alpha)+G_vx_1,\qquad x_1\to +\infty.
\eqno{(2.13)}
$$

Зеркально -- диффузное отражение ферми--частиц от поверхности означает, что последние отражаются от стенки, имея фермиевское распределение,
т.е.
$$
f(x=0,\mathbf{v})=qf_F(v)+(1-q)f(x=0,-v_x,v_y,v_z),\qquad v_x>0,
\eqno{(2.14)}
$$
где $0\leqslant q \leqslant 1$, $q$ -- коэффициент диффузности,
$f_F(v)$ -- абсолютный фермиан.

В уравнении (2.14) параметр $q$ (коэффициент диффузности) --
часть молекул, рассеивающихся границей диффузно, $1-q$ -- часть
молекул, рассеивающихся зеркально.

Учитывая, что функцию распределения мы ищем в виде (2.4), из
условия (2.14)
получаем граничное условие на стенке на функцию $h(x_1,\mu)$:
$$
h(0,\mu)=(1-q)h(0,-\mu),\qquad \mu>0.
\eqno{(2.15)}
$$

Вторым граничным условием является граничное условие "вдали от
стенки"\,. Этим условием является соотношение (2.13).
Преобразуем это условие на функцию $h(x_1,\mu)$. Условие
(2.13) означает, что вдали от стенки массовая скорость переходит
в свое асимптотическое распределение
$$
U_y^{as}(x_1)=U_{sl}(\alpha)+G_vx_1.
$$
Выражение для массовой скорости (2.8) означает, что вдали от
стенки функция $h(x_1,\mu)$ переходит в свое асимптотическое
распределение
$$
h_{as}(x_1,\mu)=2U_{sl}(\alpha)+2G_v(x_1-\mu),
$$
называемое распределением Чепмена --- Энскога (см., например,
\cite{1}--\cite{3}).

Таким образом, вторым граничным условием является условие:
$$
h(x_1,\mu)=2U_{sl}(\alpha)+2G_v(x_1-\mu),\qquad x\to+\infty.
\eqno{(2.16)}
$$

Теперь задача Крамерса при условии полного диффузного отражения
ферми--частиц от
стенки сформулирована полностью и состоит в
решении уравнения (2.12) с граничными условиями (2.15) и (2.16).
При этом требуется определить безразмерную скорость скольжения
$U_{sl}(\alpha)$, величина градиента $G_v$ считается заданной.

\begin{center}
\item{}\section{Включение граничных условий в кинетическое уравнение}
\end{center}

Продолжим функцию распределения на сопряженное полупространство
симметричным образом:
$$
f(t,x,\mathbf{v})=f(t,-x, -v_x,v_y,v_z).
\eqno{(3.1)}
$$

Продолжение согласно (3.1) на полупространство
$x<0$ позволяет включить граничные условия в уравнения задачи.

Такое продолжение функции распределения
на полупространство $x<0$
позволяет фактически рассматривать две задачи, одна из которых
определена в "положительном"\, полупространстве $x>0$, вторая -- в
отрицательном "полупространстве"\, $x<0$.

Сформулируем зеркально -- диффузные граничные условия для функции
распределения соответственно для "положительного"\, и для
"отрицательного"\, полупространств:
$$
f(t,+0, \mathbf{v})=qf_0(v)+(1-q)f(t,+0,-v_x,v_y, v_z), \quad v_x>0,
\eqno{(3.2)}
$$
$$
f(t,-0, \mathbf{v})=qf_0(v)+(1-q)f(t,-0, -v_x,v_y, v_z),
\quad v_x<0.
\eqno{(3.3)}
$$
где $q$ -- коэффициент диффузности, $0 \leqslant q \leqslant 1$.

В уравнениях (3.2) и (3.3) параметр $q$ (коэффициент диффузности) --
часть молекул,
рассеивающихся границей диффузно, $1-q$ -- часть молекул,
рассеивающихся зеркально, т.е. уходящие от стенки молекулы имеют
максвелловское распределение по скоростям.

Далее безразмерную координату $x_1$ снова будем обозначать через
$x$.

Согласно (2.4) и (3.1) мы имеем:
$$
h(x,\mu)=h(-x,-\mu), \qquad \mu>0.
$$

На функцию $h(x,\mu)$ в "положительном"\, и "отрицательном"\, полупространствах получаем одно и то же уравнение уравнение:
$$
\mu\dfrac{\partial h}{\partial x}+h(x,\mu)=
\int\limits_{-\infty}^{\infty}K(t,\alpha)h(x,t)\,dt,
\eqno{(3.4)}
$$
и соответственно следующие граничные условия:
$$
h(+0,\mu)=(1-q)h(+0,-\mu)=(1-q)h(-0,\mu), \quad \mu>0,
$$
$$
h(-0,\mu)=(1-q)h(-0,-\mu)=(1-q)h(+0,\mu), \quad \mu<0.
$$

Правая часть уравнения (3.4) есть удвоенная массовая скорость
газа:
$$
U(x)=\int\limits_{-\infty}^{\infty}K(t,\alpha)h(x,t)dt.
$$

Представим функцию $h(x,\mu)$ в виде:
$$
h(x,\mu)=h^{\pm}_{as}(x,\mu)+h_c(x,\mu),
\quad \text{если}\quad
\pm x>0,
$$
где асимптотическая часть функции распределения (так называемая
чепмен --- энскоговская функция распределения)
$$
h_{as}^{\pm}(x,\mu)=2U_{sl}(q,\alpha)\pm 2G_v(x-\mu),
\quad \text{если}\quad
\pm x>0,
\eqno{(3.5)}
$$
также является решением кинетического уравнения (3.4).

Здесь $U_{sl}(q,\alpha)$ -- есть искомая скорость изотермического
скольжения (безразмерная).

Следовательно, функция $h_c(x,\mu)$ также удовлетворяет
уравнению (3.4):
%и
%$$
%h_{as}(x,\mu)=2U_0-2g_v(x-\mu), \quad \text{если}\quad x<0.
%$$
$$
\mu\dfrac{\partial h_c}{\partial x}+h_c(x,\mu)=
\int\limits_{-\infty}^{\infty}K(t,\alpha)h_c(x,t)dt.
$$

Так как вдали от стенки ($x\to \pm \infty$) функция распределения
$h(x,\mu)$ переходит в чепмен --- энскоговскую
$h_{as}^{\pm}(x,\mu)$, то для функции $h_c(x,\mu)$, отвечающей
непрерывному спектру, получаем следующее граничное условие:
$h_c(\pm \infty,\mu)=0.$

Отсюда для массовой скорости газа получаем:
$$
U_c(\pm \infty)=0.
\eqno{(3.6)}
$$

Отметим, что в равенстве (3.5) знак градиента в "отрицательном"\,
полупространстве меняется на противоположный. Поэтому условие (3.6)
выполняется автоматически для функций $h_{as}^{\pm}(x,\mu)$.

Тогда граничные условия переходят в следующие:
$$
h_c(+0,\mu)=$$$$=-h_{as}^+(+0,\mu)+(1-q)h_{as}^+(+0,-\mu)+%$$$$+
(1-q)h_c(+0,-\mu),
\quad \mu>0,
$$
$$
h_c(-0,\mu)=$$$$=-h_{as}^-(-0,\mu)+(1-q)h_{as}^-(-0,-\mu)%+$$$$+
(1-q)h_c(-0,-\mu), \quad \mu<0.
$$

Обозначим
$$
h_0^{\pm}(\mu)=-2qU_{sl}(q,\alpha)+(2-q)2G_v|\mu|.
$$

И перепишем предыдущие граничные условия в виде:
$$
h_c(+0,\mu)=h_0^+(\mu)+(1-q)h_c(+0,-\mu), \quad \mu>0,
$$
$$
h_c(-0,\mu)=h_0^-(\mu)+(1-q)h_c(-0,-\mu), \quad \mu<0,
$$
где
$$
h_0^{\pm}(\mu)=-h_{as}^{\pm}(0,\mu)+(1-q)h_{as}^{\pm}(0,-\mu)=
$$$$=-2qU_{sl}(q,\alpha)+(2-q)2G_v|\mu|.
$$

Учитывая симметричное продолжение функции распределения, имеем
$$ h_c(-0,-\mu)=h_c(+0,+\mu),\qquad
h_c(+0,-\mu)=h_c(-0,+\mu).
$$
Следовательно, граничные условия перепишутся в виде:
$$
h_c(+0,\mu)=h_0^+(\mu)+(1-q)h_c(-0,\mu), \quad \mu>0,
\eqno{(3.7)}
$$
$$
h_c(-0,\mu)=h_0^-(\mu)+(1-q)h_c(+0,\mu), \quad \mu<0.
\eqno{(3.8)}
$$

Включим граничные условия (3.7) и (3.8) в кинетическое
уравнение следующим образом:
$$
\mu \dfrac{\partial h_c}{\partial x}+h_c(x,\mu)=2U_c(x)+
|\mu|\Big[h_0^{\pm}(\mu)-q h_c(\pm 0,\mu)\Big]
\delta(x),
\eqno{(3.9)}
$$
где $U_c(x)$ -- часть массовой скорости, отвечающая непрерывному
спектру,
$$
2U_c(x)=\int\limits_{-\infty}^{\infty}K(t,\alpha)h_c(x,t)\,dt.
\eqno{(3.10)}
$$

Уравнение (3.9) содержит два уравнения. В "положительном"\, полупространстве,
т.е. при $x>0$ в правой части уравнения (3.9) следует взять
верхний знак "плюс"\,, а в "нижнем"\, полупространстве, т.е. при $x<0$
в правой части того же уравнения следует взять знак "минус"\,.

В самом деле, пусть, например, $\mu>0$.
Проинтегрируем обе части уравнения
(3.9) по $x$ от $-\varepsilon$ до $+\varepsilon$. В результате получаем
равенство:
$$
h_c(+\varepsilon,\mu)-h_c(-\varepsilon,\mu)=h_0^+(\mu)-
qh_c(-\varepsilon,\mu),
%\eqno{(1.13)}
$$
откуда переходя к пределу при $\varepsilon\to 0$
в точности получаем граничное условие (3.7).

На основании определения массовой скорости (3.10) заключаем, что
для нее выполняется условие (3.6):
$U_c(+\infty)=0.$
Следовательно, в полупространстве $x>0$ профиль массовой скорости газа
вычисляется по формуле:
$$
U(x)=U_{as}(x)+\dfrac{1}{2}\int\limits_{-\infty}^{\infty}
K(t,\alpha)h_c(x,t)dt,
\eqno{(3.11)}
$$
а вдали от стенки имеет следующее линейное распределение:
$$
U_{as}(x)=U_{sl}(q,\alpha)+G_vx, \qquad x\to +\infty.
\eqno{(3.12)}
$$

\begin{center}
\item{}\section{Кинетическое уравнение во втором и четвертом
квадрантах фазового пространства}
\end{center}

Решая уравнение (3.9) при $x>0,\,\mu<0$, считая заданным массовую
скорость $U(x)$, получаем, удовлетворяя граничным условиям (3.8),
следующее решение:
$$
h_c^+(x,\mu)=-\dfrac{1}{\mu}\exp(-\dfrac{x}{\mu})
\int\limits_{x}^{+\infty} \exp(+\dfrac{t}{\mu})2U_c(t)\,dt.
\eqno{(4.1)}
$$

Аналогично при $x<0,\,\mu>0$ находим:
$$
h_c^-(x,\mu)=-\dfrac{1}{\mu}\exp(-\dfrac{x}{\mu})
\int\limits_{x}^{-\infty} \exp(+\dfrac{t}{\mu})2U_c(t)\,dt.
\eqno{(4.2)}
$$

Теперь уравнения (3.9) и (3.10) можно переписать, заменив второй член в
квадратной скобке из (3.9) согласно (4.1) и (4.2), в виде:
$$
\mu\dfrac{\partial h_c}{\partial x}+h_c(x,\mu)=2U_c(x)+
|\mu|\Big[h_0^{\pm}(\mu)-qh_c^{\pm}(0,\mu)\Big]\delta(x),
\eqno{(4.3)}
$$
$$
2U_c(x)=\int\limits_{-\infty}^{\infty}K(t,\alpha)h_c(x,t)dt.
\eqno{(4.4)}
$$

В равенствах (4.3) граничные значения $h_c^{\pm}(0,\mu)$
выражаются через составляющую массовой скорости, отвечающей
непрерывному спектру:
$$
h_{c}^{\pm}(0,\mu)=-\dfrac{1}{\mu}e^{-x/\mu} \int\limits_{0}^{\pm
\infty}e^{t/\mu}2U_c(t)dt=h_c(\pm 0,\mu).
$$

Решение уравнений (4.4) и (4.3) ищем в виде интегралов Фурье:
$$
2U_c(x)=\dfrac{1}{2\pi}\int\limits_{-\infty}^{\infty}
e^{ikx}E(k)\,dk,\qquad
\delta(x)=\dfrac{1}{2\pi}\int\limits_{-\infty}^{\infty}
e^{ikx}\,dk,
\eqno{(4.5)}
$$
$$
h_c(x,\mu)=\dfrac{1}{2\pi}\int\limits_{-\infty}^{\infty}
e^{ikx}\Phi(k,\mu)\,dk.
\eqno{(4.6)}
$$

При этом функция распределения $h_c^+(x,\mu)$ выражается через
спектральную плотность $E(k)$ массовой скорости следующим образом:
$$
h_c^+(x,\mu)=-\dfrac{1}{\mu}\exp(-\dfrac{x}{\mu})
\int\limits_{x}^{+\infty} \exp(+\dfrac{t}{\mu})dt
\dfrac{1}{2\pi}
\int\limits_{-\infty}^{+\infty}e^{ikt}E(k,\mu)\,dk=
$$
$$
=\dfrac{1}{2\pi}\int\limits_{-\infty}^{\infty}\dfrac{e^{ikx}
E(k,\mu)}{1+ik\mu}dk.
%\eqno{(2.7)}
$$

Аналогично,

$$
h_c^-(x,\mu)=\dfrac{1}{2\pi}
\int\limits_{-\infty}^{\infty}\dfrac{e^{ikx}
E(k,\mu)}{1+ik\mu}dk.
%\eqno{(2.8)}
$$

Таким образом,
$$
h_c^{\pm}(x,\mu)=\dfrac{1}{2\pi}
\int\limits_{-\infty}^{\infty}\dfrac{e^{ikx}
E(k,\mu)}{1+ik\mu}dk.
%\eqno{(2.8)}
$$

Используя четность функции $E(k)$ далее получаем:
$$
h_c^{\pm}(0,\mu)=\dfrac{1}{2\pi}
\int\limits_{-\infty}^{\infty}\dfrac{E(k,\mu)}{1+ik\mu}dk=
\dfrac{1}{2\pi}\int\limits_{-\infty}^{\infty}
\dfrac{E(k)\,dk}{1+k^2\mu^2}=%$$$$=
\dfrac{1}{\pi}\int\limits_{0}^{\infty}\dfrac{E(k)\,dk}{1+
k^2\mu^2}.
\eqno{(4.7)}
$$

Теперь с помощью равенства (4.7) уравнение (4.3) можно переписать в виде:
$$
\mu\dfrac{\partial h_c}{\partial x}+h_c(x,\mu)=2U_c(x)+
|\mu|\Bigg[h_0^{\pm}(\mu)-\dfrac{q}{\pi}\int\limits_{0}^{\infty}
\dfrac{E(k)\,dk}{1+k^2\mu^2}\Bigg]\delta(x),
\eqno{(4.3')}
$$

\begin{center}
\item{}\section{Характеристическая система}
\end{center}

Теперь подставим интегралы Фурье (4.6) и (4.5), а также равенство
(4.7) в уравнения (4.3) и (4.4). Получаем характеристическую систему
уравнений:
$$
\Phi(k,\mu)(1+ik\mu)=$$$$=E(k)+|\mu|\Bigg[-2qU_{sl}(q,\alpha)+2(2-q)G_v|\mu|
-\dfrac{q}{\pi}
\int\limits_{0}^{\infty}\dfrac{E(k_1)dk_1}{1+k_1^2\mu^2}\Bigg],
\eqno{(5.1)}
$$
$$
E(k)=\int\limits_{-\infty}^{\infty}K(t,\alpha)\Phi(k,t)dt.
\eqno{(5.2)}
$$

Из уравнения (5.1) получаем:
$$
\Phi(k,\mu)=\dfrac{E(k)}{1+ik\mu}+$$$$+
\dfrac{|\mu|}{1+ik\mu}\Bigg[-2qU_{sl}(q,\alpha)+2(2-q)G_v|\mu|-\dfrac{q}{\pi}
\int\limits_{0}^{\infty}\dfrac{E(k_1)dk_1}{1+k_1^2\mu^2}\Bigg],
\eqno{(5.3)}
$$

Подставим выражение для функции $\Phi(k,\mu)$, определенное равенством
(5.3), в (5.2).
Получаем, что:
$$
E(k)L(k)=-2qU_{sl}(q,\alpha)T_1(k)+2(2-q)G_v T_2(k)-
$$
$$
-\dfrac{q}{\pi}\int\limits_{0}^{\infty}
E(k_1)dk_1\int\limits_{-\infty}^{\infty}
\dfrac{K(t,\alpha)|t|dt}{(1+ikt)(1+k_1^2t^2)}.
\eqno{(5.4)}
$$
Здесь
$$
T_n(k)=2\int\limits_{0}^{\infty}
\dfrac{K(t,\alpha)t^n\,dt}{1+k^2t^2},\quad n=0,1,2,3\cdots,
$$
причем для четных $n$
$$
T_n(k)=2\int\limits_{0}^{\infty}
\dfrac{K(t,\alpha)t^n\,dt}{1+k^2t^2}=
\int\limits_{-\infty}^{\infty}
\dfrac{K(t,\alpha)t^n\,dt}{1+k^2t^2},\quad n=0,2,4,\cdots,
$$
кроме того,
$$
L(k)=1-\int\limits_{-\infty}^{\infty}
\dfrac{K(t,\alpha)dt}{1+ikt}.
$$
Нетрудно видеть, что
$$
L(k)=1-
\int\limits_{-\infty}^{\infty}\dfrac{K(t,\alpha)dt}{1+k^2t^2}=$$$$=
1-2\int\limits_{0}^{\infty}\dfrac{K(t,\alpha)dt}
{1+k^2t^2}=2k^2 \int\limits_{0}^{\infty}
\dfrac{K(t,\alpha)t^2\;dt}{1+k^2t^2}=k^2 \int\limits_{-\infty}^{\infty}
\dfrac{K(t,\alpha)t^2\;dt}{1+k^2t^2},
$$
или, кратко,
$$
L(k)=k^2 T_2(k).
$$

Кроме того, внутренний интеграл в (5.4) преобразуем
и обозначим следующим образом:
$$
\int\limits_{-\infty}^{\infty}
\dfrac{K(t,\alpha)|t|dt}{(1+ikt)(1+k_1^2t^2)}
=2\int\limits_{0}^{\infty}\dfrac{K(t,\alpha)t\,dt}
{(1+k^2t^2)(1+k_1^2t^2)}=J(k,k_1).
$$

Заметим, что
$$
J(k,0)=T_1(k), \qquad J(0,k_1)=T_1(k_1).
$$

Перепишем теперь уравнение (5.4) с помощью предыдущего равенства в
следующем виде:
$$
E(k)L(k)=-2qU_{sl}(q,\alpha)T_1(k)+2(2-q)G_v T_2(k)-$$$$-
\dfrac{q}{\pi}\int\limits_{0}^{\infty} J(k,k_1)E(k_1)\,dk_1.
\eqno{(5.5)}
$$

Уравнение (5.5) есть интегральное уравнение Фредгольма второго
рода.

\begin{center}
\item{}\section{Ряд Неймана}
\end{center}

Считая градиент массовой скорости в %характеристическом
уравнении (5.5) заданным, разложим решения характеристической
системы (5.3) и (5.5) в ряд по степеням
коэффициента диффузности $q$:
$$
E(k)=G_v2(2-q)\Big[E_0(k)+q\,E_1(k)+q^2\,E_2(k)+\cdots\big],
\eqno{(6.1)}
$$
$$
\Phi(k,\mu)=G_v2(2-q)\Big[
\Phi_0(k,\mu)+q\Phi_1(k,\mu)+q^2\Phi_2(k,\mu)+\cdots\Big].
\eqno{(6.2)}
$$

Скорость скольжения $U_{sl}(q,\alpha)$ при этом будем искать в виде
$$
U_{sl}(q,\alpha)=G_v\dfrac{2-q}{q}
\Big[U_0+U_1q+U_2q^2+\cdots+U_nq^n+\cdots\Big].
\eqno{(6.3)}
$$

Подставим ряды (6.1)--(6.3) в уравнения (5.3) и (5.5). Получаем
следующую систему уравнений:
$$
(1+ik\mu)[\Phi_0(k,\mu)+\Phi_1(k,\mu)q+\Phi_2(k,\mu)q^2+\cdots]=
$$
$$
=[E_0(k)+E_1(k)q+E_2(k)q^2+\cdots]-(U_0+U_1q+U_2q^2+\cdots)|\mu|+
$$
$$
+\mu^2-|\mu|\dfrac{q}{\pi}\int\limits_{0}^{\infty}
\dfrac{E_0(k_1)+E_1(k_1)q+E_2(k_1)q^2+\cdots}{1+k_1^2\mu^2}dk_1,
$$
$$
[E_0(k)+E_1(k)q+E_2(k)q^2+\cdots]L(k)=-[U_0+U_1q+U_2q^2+\cdots]T_1(k)+
T_2(k)-
$$
$$
-\dfrac{q}{\pi}\int\limits_{0}^{\infty}J(k,k_1)
[E_0(k_1)+E_1(k_1)q+E_2(k_1)q^2+\cdots]dk_1.
$$

Последние интегральные уравнения распадаются на
эквивалентную бесконечную систему уравнений.
В нулевом приближении получаем следующую систему уравнений:
$$
E_0(k)L(k)=T_2(k)-U_0T_1(k),
\eqno{(6.4)}
$$
$$
\Phi_0(k,\mu)(1+ik\mu)=E_0(k)+\mu^2-U_0|\mu|,
\eqno{(6.5)}
$$

В первом приближении:
$$
E_1(k)L(k)=-U_1T_1(k)-
\dfrac{1}{\pi}\int\limits_{0}^{\infty}
J(k,k_1)E_0(k_1)dk_1,
\eqno{(6.6)}
$$
$$
\Phi_1(k,\mu)(1+ik\mu)=E_1(k)-U_1|\mu|-\dfrac{|\mu|}{\pi}
\int\limits_{0}^{\infty}\dfrac{E_0(k_1)dk_1}{1+k_1^2\mu^2}.
\eqno{(6.7)}
$$

Во втором приближении:
$$
E_2(k)L(k)=-U_2T_1(k)-\dfrac{1}{\pi}
\int\limits_{0}^{\infty}J(k,k_2)E_1(k_2)\,dk_2,
\eqno{(6.8)}
$$
$$
\Phi_2(k,\mu)(1+ik\mu)=E_2(k)-U_2|\mu|-
\dfrac{|\mu|}{\pi}\int\limits_{0}^{\infty}\dfrac{E_1(k_2)dk_2}
{1+k_2^2\mu^2}.
\eqno{(6.9)}
$$

В $n$--м приближении получаем:
$$
E_n(k)L(k)=-U_nT_1(k)-\dfrac{1}{\pi}
\int\limits_{0}^{\infty}J(k,k_n)E_{n-1}(k_n)dk_n,
\eqno{(6.10)}
$$
$$
\Phi_n(k,\mu)(1+ik\mu)=E_n(k)-U_n|\mu|-\hspace{5cm}
$$
$$
\hspace{4.5cm}
- \dfrac{|\mu|}{\pi}
\int\limits_{0}^{\infty}\dfrac{E_{n-1}(k_n)dk_n}{1+k_n^2\mu^2},
\quad n=1,2,3,\cdots.
\eqno{(6.11)}
$$

\begin{center}
  \item{}\subsection{Нулевое приближение}
\end{center}

Из формулы (6.4) для нулевого приближения находим:
$$
E_0(k)=\dfrac{T_2(k)-U_0T_1(k)}{L(k)}.
\eqno{(6.12)}
$$

Нулевое приближение массовой скорости на основании (6.12) равно:
$$
U_c^{(0)}(x)=G_v\dfrac{2-q}{2\pi}\int\limits_{-\infty}^{\infty}
e^{ikx}E_0(k)\,dk=$$$$=G_v\dfrac{2-q}{2\pi}
\int\limits_{-\infty}^{\infty}
e^{ikx}\dfrac{-U_0T_1(k)+T_2(k)}{L(k)}dk.
\eqno{(6.13)}
$$

Согласно (6.13) наложим на нулевое приближение массовой скорости
требование: $U_c(+\infty)=0$. Это условие приводит к тому, что
подынтегральное выражение из интеграла Фурье (6.13) в точке $k=0$
конечно. Следовательно, мы должны устранить полюс второго
порядка в точке $k=0$ у функции $E_0(k)$.

Замечая, что
$$
T_2(0)=\int\limits_{-\infty}^{\infty}t^2K(t,\alpha)dt=
\dfrac{1}{2l_0^F(\alpha)}\int\limits_{-\infty}^{\infty}
t^2\ln(1+e^{\alpha-t^2})dt=\dfrac{l_2^F(\alpha)}{l_0^F(\alpha)}.
$$
$$
T_1(0)=2\int\limits_{0}^{\infty}tK(t,\alpha)dt=
\dfrac{1}{l_0^F(\alpha)}\int\limits_{0}^{\infty}t\ln(1+e^{\alpha-t^2})dt=
\dfrac{l_1^F(\alpha)}{l_0^F(\alpha)},
$$
находим нулевое приближение $U_0$:
$$
U_0=\dfrac{T_2(0)}{T_1(0)}=\dfrac{\int\limits_{-\infty}^{\infty}
t^2\ln(1+e^{\alpha-t^2})dt}
{2\int\limits_{0}^{\infty}t\ln(1+e^{\alpha-t^2})dt}=
\dfrac{l_2^F(\alpha)}{l_1^F(\alpha)}.
%\eqno{(4.13)}
$$

Заметим, что
$$
U_0(-\infty)=\dfrac{\sqrt{\pi}}{2}=0.8862.
$$

Найдем числитель выражения (6.12):
$$
T_2(k)-U_0T_1(k)=T_2(k)-\dfrac{T_2(0)}{T_1(0)}T_1(k)=$$$$=
\dfrac{1}{T_1(0)}\Big[T_1(0)T_2(k)-T_2(0)T_1(k)\Big].
$$

Замечая, что
$$
\dfrac{1}{1+k^2t^2}=1-\dfrac{k^2t^2}{1+k^2t^2},
$$
получаем
$$
T_2(k)=T_2(0)-k^2T_4(k),\qquad T_1(k)=T_1(0)-k^2T_3(k).
$$

Откуда
$$
T_1(0)T_2(k)-T_2(0)T_1(k)=k^2\Big[T_2(0)T_3(k)-T_1(0)T_4(k)\Big].
$$

Следовательно, мы получаем:
$$
T_2(k)-U_0T_1(k)=\dfrac{k^2}{L(k)T_1(0)}\Big[T_2(0)T_3(k)-T_1(0)T_4(k)\Big],
$$
или, учитывая, что $L(k)=k^2T_2(k)$, запишем предыдущее равенство короче,
$$
E_0(k)=\dfrac{\varphi_0(k)}{T_2(k)},
$$
где
$$
\varphi_0(k)=\dfrac{T_2(0)T_3(k)-T_1(0)T_4(k)}{T_1(0)}.
$$

Согласно (6.5) находим:
$$
\Phi_0(k,\mu)= \dfrac{E_0(k)+\mu^2-U_0|\mu|}
{1+ik\mu},
%\eqno{(4.15)}
$$
и, следовательно,
$$
h_c^{(0)}(x,\mu)=\dfrac{1}{2\pi}\int\limits_{-\infty}^{\infty}
\Big[E_0(k)+\mu^2-U_0|\mu|\Big]
\dfrac{e^{ikx}dk}{1+ik\mu}.
$$

\begin{center}
  \item{}\subsection{Первое приближение}
\end{center}

Перейдем к первому приближению. В первом приближении из уравнения
(6.6) находим:
$$
E_1(k)=-\dfrac{1}{L(k)}\Big[U_1T_1(k)+\dfrac{1}{\pi}
\int\limits_{0}^{\infty}
\dfrac{J(k,k_1)}{T_2(k_1)}\varphi_0(k_1)dk_1\Big].
\eqno{(6.14)}
$$

Первая поправка к массовой скорости имеет вид
$$
U_c^{(1)}(x)=G_v\dfrac{2-q}{2\pi}\int\limits_{-\infty}^{\infty}
e^{ikx}E_1(k)\,dk.
%\eqno{()}
$$

Требование $U_c(+\infty)=0$ приводит к требованию конечности
подынтегрального выражения в предыдущем интеграле Фурье.
Устраняя полюс второго порядка в точке $k=0$, находим:
$$
U_1=-
\dfrac{1}{\pi T_1(0)}\int\limits_{0}^{\infty}
J(0,k_1)\dfrac{\varphi_0(k_1)}{T_2(k_1)}dk_1=
$$
$$
=-\dfrac{1}{\pi T_1(0)}
\int\limits_{0}^{\infty}\dfrac{T_1(k_1)}{T_2(k_1)}\varphi_0(k_1)\,dk_1=
-\dfrac{1}{\pi T_1(0)}
\int\limits_{0}^{\infty}T_1(k_1)E_0(k_1)\,dk_1.
\eqno{(6.15)}
$$

Нетрудно проверить, что
$$
U_1(-\infty)\approx 0.1405.
$$

Преобразуем с помощью (6.15) выражение в квадратной
скобке из выражения (6.14):
$$
U_1T_1(k)+\dfrac{1}{\pi}
\int\limits_{0}^{\infty}J(k,k_1)\dfrac{\varphi_0(k_1)}{T_2(k_1)}dk_1=
$$

$$
=\dfrac{1}{\pi}\int\limits_{0}^{\infty}
J(k,k_1)\dfrac{\varphi_0(k_1)}{T_2(k_1)}dk_1-
\dfrac{T_1(k)}{\pi T_1(0)}\int\limits_{0}^{\infty}
T_1(k_1)\dfrac{\varphi_0(k_1)}{T_2(k_1)}dk_1=
$$
$$
=\dfrac{1}{\pi}\int\limits_{0}^{\infty}
\Big[J(k,k_1)-\dfrac{T_1(k)T_1(k_1)}{T_1(0)}\Big]E_0(k_1)dk_1.
\eqno{(4.16)}
$$

Заметим, что $J(0,k_1)=T_1(k_1)$. Найдем выражение
$$
J(k,k_1)-\dfrac{T_1(k)T_1(k_1)}{T_1(0)}.
$$
Рассмотрим разложение на элементарные дроби:
$$
\dfrac{1}{(1+k^2t^2)(1+k_1^2t^2)}=
\dfrac{[(1+k_1^2t^2)-k_1^2t^2][(1+k^2t^2)-k^2t^2]}
{(1+k^2t^2)(1+k_1^2t^2)}=
$$
$$
=1-\dfrac{k_1^2t^2}{1+k_1^2t^2}-\dfrac{k^2t^2}{1+k^2t^2}+
\dfrac{k^2k_1^2\,t^4}{(1+k^2t^2)(1+k_1^2t^2)}
$$

С помощью этого разложения преобразуем интеграл
$$
J(k,k_1)=2\int\limits_{0}^{\infty}\dfrac{K(t,\alpha)t\,dt}
{(1+k^2t^2)(1+k_1^2t^2)}.
$$

Получаем следующее представление этого интеграла:
$$
J(k,k_1)=T_1(0)-k_1^22\int\limits_{0}^{\infty}\dfrac{K(t,\alpha)t^3dt}
{1+k_1^2t^2}-k^2 2\int\limits_{0}^{\infty}
\dfrac{K(t,\alpha)t^3dt}{1+k^2t^2}+$$$$+k^2k_1^22
\int\limits_{0}^{\infty}\dfrac{K(t,\alpha)t^5dt}{(1+k^2t^2)(1+k_1^2t^2)},
$$
или
$$
J(k,k_1)=T_1(0)-k^2T_3(k)-k_1^2T_3(k_1)+k^2k_1^2J_5(k,k_1),
$$
где
$$
J_n(k,k_1)=2\int\limits_{0}^{\infty}
\dfrac{K(t,\alpha)t^ndt}{(1+k^2t^2)(1+k_1^2t^2)}, \qquad n=3,5.
$$

Теперь ясно, что
$$
J(k,k_1)-\dfrac{T_1(k)T_1(k_1)}{T_1(0)}=
k^2k_1^2\Big[J_5(k,k_1)-\dfrac{T_3(k)T_3(k_1)}{T_1(0)}\Big].
$$
Представим это выражение в виде
$$
J(k,k_1)-\dfrac{T_1(k)T_1(k_1)}{T_1(0)}=k^2 S(k,k_1),
$$
где
$$
S(k,k_1)=k_1^2\Big[J_5(k,k_1)- \dfrac{T_3(k)T_3(k_1)}{T_1(0)}\Big].
$$

Вернемся к выражению (6.14). С помощью (6.16) теперь получаем:
$$
E_1(k_1)=-\dfrac{1}{\pi T_2(k_1)}
\int\limits_{0}^{\infty}
\dfrac{S(k_1,k_2)}{T_2(k_2)}\varphi_0(k_2)\,dk_2,
\eqno{(6.17)}
$$
или, кратко,
$$
E_1(k_1)=\dfrac{\varphi_1(k_1)}{T_2(k_1)},
$$
где
$$
\varphi_1(k_1)=-\dfrac{1}{\pi}\int\limits_{0}^{\infty}
\dfrac{S(k_1,k_2)}{T_2(k_2)}\varphi_0(k_2)\,dk_2,
$$
или
$$
\varphi_1(k_1)=-\dfrac{1}{\pi}\int\limits_{0}^{\infty}
S(k_1,k_2)E_0(k_2)\,dk_2.
$$

Теперь подставляя (6.17) в (6.7) находим первое приближение
спектральной плотности функции распределения:
$$
\Phi_1(k,\mu)=\dfrac{1}{1+ik\mu}\Big[E_1(k)
-U_1|\mu|-\dfrac{|\mu|}{\pi}
\int\limits_{0}^{\infty}\dfrac{E_0(k_1)\,dk_1}
{1+k_1^2\mu^2}\Big].
%\eqno{(3.24)}
$$
\begin{center}
  \item{}\subsection{Второе приближение}
\end{center}

Перейдем ко второму приближению задачи -- уравнения (6.8) и (6.9).
Из уравнения (6.8) находим:
$$
E_2(k)=-\dfrac{1}{L(k)}\Big[U_2T_1(k)+\dfrac{1}{\pi}
\int\limits_{0}^{\infty}J(k,k_1)E_1(k_1)\,dk_1\Big].
\eqno{(6.18)}
$$

Вторая поправка к массовой скорости имеет вид:
$$
U_c^{(2)}(x)=G_v\dfrac{2-q}{2\pi}\int\limits_{-\infty}^{\infty}
e^{ikx}E_2(k)\,dk.
$$

Условие $U_c(+\infty)=0$ приводит к требованию ограниченности
функции $E_2(k)$ в точке $k=0$.
Устраняя полюс второго порядка в точке $k=0$ в правой части
равенства для $E_2(k)$, находим:
$$
U_2=-\dfrac{1}{\pi T_1(0)}\int\limits_{0}^{\infty}J(0,k_1)E_1(k_1)dk_1=$$$$=
-\dfrac{1}{\pi T_1(0)}\int\limits_{0}^{\infty}T_1(k_1)E_1(k_1)dk_1.
\eqno{(6.19)}
$$

Нетрудно проверить, что
$$
U_2(-\infty)\approx -0.0116.
$$

Формулу (6.19) преобразуем к следующему виду:
$$
U_2=\dfrac{1}{\pi^2T_1(0)}\int\limits_{0}^{\infty}\int\limits_{0}^{\infty}
\dfrac{T_1(k_1)S(k_1,k_2)}{T_2(k_1)T_2(k_2)}\varphi_0(k_2)\,
dk_1dk_2.
$$

Преобразуем выражение (6.18) с помощью равенства (6.19). Имеем:
$$
E_2(k)=-\dfrac{1}{L(k)}
\int\limits_{0}^{\infty}\Big[J(k,k_1)-\dfrac{T_1(k)T_1(k_1)}{T_1(0)}\Big]
E_1(k_1)\,dk_1.
$$

Выше было показано, что
$$
J(k,k_1)-\dfrac{T_1(k)T_1(k_1)}{T_1(0)}= k^2 S(k,k_1).
$$

Следовательно, предыдущее равенство дает:
$$
E_2(k)=-\dfrac{1}{\pi T_2(k)}
\int\limits_{0}^{\infty}S(k,k_1)E_1(k_1)\,dk_1=$$$$=
\dfrac{1}{\pi^2T_2(k)}\int\limits_{0}^{\infty}
\int\limits_{0}^{\infty} \dfrac{S(k,k_1)S(k_1,k_2)}{T_2(k_1)T_2(k_2)}
\varphi_0(k_2)dk_1dk_2.
$$

Перепишем это равенство в виде:
$$
E_2(k)=\dfrac{\varphi_2(k)}{T_2(k)},
$$
где
$$
\varphi_2(k)=-\dfrac{1}{\pi}\int\limits_{0}^{\infty}
S(k,k_1)E_1(k_1)dk_1=
$$
$$
=\dfrac{1}{\pi^2}\int\limits_{0}^{\infty}\int\limits_{0}^{\infty}
\dfrac{S(k,k_1)S(k_1,k_2)}{T_2(k_1)T_2(k_2)}
\varphi_0(k_2)dk_1dk_2.
$$

Для второго приближения спектральной плотности функции распределения
из уравнения (6.9) получаем:
$$
\Phi_2(k,\mu)=\dfrac{1}{1+ik\mu}\Bigg[E_2(k)-U_2|\mu|-
-\dfrac{|\mu|}{\pi}
\int\limits_{0}^{\infty}\dfrac{E_1(k_1)dk_1}
{1+k_1^2\mu^2}\Bigg].
$$

\begin{center}
  \item{}\subsection{Высшие приближения}
\end{center}

В третьем приближении получаем:
$$
E_3(k)=-\dfrac{1}{L(k)}\Big[U_3T_1(k)+\dfrac{1}{\pi}
\int\limits_{0}^{\infty}J(k,k_1)E_2(k_1)dk_1\Big].
$$
Как и ранее, устраняя полюс второго порядка в точке $k=0$, получаем:
$$
U_3=-\dfrac{1}{\pi T_1(0)}\int\limits_{0}^{\infty}J(0,k_1)E_2(k_1)dk_1=
-\dfrac{1}{\pi T_1(0)}\int\limits_{0}^{\infty}T_1(k_1)E_2(k_1)dk_1,
$$
или
$$
U_3=-\dfrac{1}{\pi T_1(0)}\int\limits_{0}^{\infty}
\dfrac{T_1(k_1)}{T_2(k_1)}\varphi_2(k_1)\,dk_1.
$$

Заметим, что
$$
U_3(-\infty)=0.0011.
$$

Кроме того, в третьем приближении мы получаем:
$$
E_3(k)=-\dfrac{1}{\pi L(k)}\int\limits_{0}^{\infty}\Big[J(k,k_1)-
\dfrac{T_1(k)T_1(k_1)}{T_1(0)}\Big]E_2(k_1)dk_1,=
$$
$$
=-\dfrac{1}{\pi
T_2(k)}\int\limits_{0}^{\infty}S(k,k_1)E_2(k_1)dk_1,
$$
или
$$
E_3(k)=\dfrac{\varphi_3(k)}{T_2(k)},
$$
где
$$
\varphi_3(k)=-\dfrac{1}{\pi}\int\limits_{0}^{\infty}
S(k,k_1)E_2(k_1)dk_1=
$$
$$
=-\dfrac{1}{\pi^3}\int\limits_{0}^{\infty}\int\limits_{0}^{\infty}
\int\limits_{0}^{\infty}\dfrac{S(k,k_1)S(k_1,k_2)S(k_2,k_3)}
{T_2(k_1)T_2(k_2)T_2(k_3)}\varphi_0(k_3)dk_1dk_2dk_3,
$$
и
$$
U_3=-\dfrac{1}{\pi^3}\int\limits_{0}^{\infty}\int\limits_{0}^{\infty}
\int\limits_{0}^{\infty}\dfrac{T_1(k_1)S(k_1,k_2)S(k_2,k_3)}
{T_1(0)T_2(k_1)T_2(k_2)T_3(k_3)}\varphi_0(k_3)dk_1dk_2dk_3.
$$

Проводя аналогичные рассуждения, для $n$--го приближения
согласно (4.10) и (4.11) получаем:
$$
U_n=-\dfrac{1}{\pi T_1(0)}\int\limits_{0}^{\infty}T_1(k)E_{n-1}(k)\,dk,
\qquad n=1,2,\cdots
$$
$$
E_n(k)=-\dfrac{1}{\pi T_2(k)}\int\limits_{0}^{\infty}
S(k,k_1)E_{n-1}(k_1)dk_1, \qquad n=1,2,\cdots,
$$
или
$$
E_n(k)=\dfrac{\varphi_n(k)}{T_2(k)}, \qquad n=0,1,2, \cdots,
$$
где
$$
\varphi_{n}(k)=-\dfrac{1}{\pi}
\int\limits_{0}^{\infty}S(k,k_1)E_{n-1}(k_1)dk_1, \qquad
n=1,2,\cdots,
$$
$$
\Phi_n(k,\mu)=\dfrac{1}{1+ik\mu}\Bigg[E_n(k)-U_n|\mu|-
\dfrac{|\mu|}{\pi}
\int\limits_{0}^{\infty}\dfrac{E_{n-1}(k_1)dk_1}
{1+k_1^2\mu^2}\Bigg].
$$

Выпишем $n$--ые приближения $V_n$, $E_n(k)$ и $\varphi_n(k)$,
выраженные через
нулевое приближение спектральной плотности массовой скорости
$E_0(k)=\varphi_0(k)/T_2(k)$. Имеем:
$$
U_n=\dfrac{(-1)^n}{\pi^n}\int\limits_{0}^{\infty}\cdots
\int\limits_{0}^{\infty}\dfrac{T_1(k_1)S(k_1,k_2)\cdots
S(k_{n-1},k_n)}{T_1(0)T_2(k_1)\cdots T_2(k_n)}\times
$$
$$
\times\varphi_0(k_n)\,dk_1
\cdots dk_n,
$$
$$
E_n(k)=\dfrac{(-1)^n}{\pi^n T_2(k)}\int\limits_{0}^{\infty}
\cdots \int\limits_{0}^{\infty}\dfrac{S(k,k_1)S(k_1,k_2)\cdots
S(k_{n-1},k_n)}{T_2(k_1)\cdots T_2(k_n)} \times $$$$
\times\varphi_0(k_n)dk_1
\cdots dk_n,\qquad
n=1,2,3,\cdots,
$$
$$
\varphi_n(k)=\dfrac{(-1)^n}{\pi^{n}}\int\limits_{0}^{\infty}\cdots
\int\limits_{0}^{\infty}\dfrac{S(k,k_1)S(k_1,k_2)\cdots S(k_{n-1},k_n)}
{T_2(k_1)\cdots T_2(k_n)}\times
$$
$$
\times \varphi_0(k_n)dk_1\cdots dk_n,\qquad n=1,2,\cdots.
$$
\begin{center}
  \item{}\section{Сравнение с точным решением. Скорость скольжения}
\end{center}

Сравним нулевое, первое, второе и третье приближения
при $q=1$ с точным решением.
Ограничимся случаем квантовых ферми--газов, близких к классическим
(т.е. при $\alpha\to -\infty$), и случаем диффузного отражения молекул
газа от поверхности.

Точное значение скорости скольжения в случае диффузного рассеяния для
квантовых ферми--газов с постоянной частотой столкновений таково:
$$
U_{sl}(q=1,\alpha)=V_1(\alpha)G_v.
$$

Здесь
$$
V_1(\alpha)=-\dfrac{1}{\pi}\int\limits_{0}^{\infty}\zeta(\tau,\alpha)d\tau,
$$
$$
\zeta(\tau,\alpha)=\theta(\tau,\alpha)-\pi,\qquad
\theta(\tau,\alpha)=\arg \lambda^+(\mu,\alpha)=
\arctg\dfrac{\lambda(\tau,\alpha)}{\pi \tau K(\tau,\alpha)},
$$
или, в явном виде,
$$
\theta(\tau,\alpha)=\arcctg\Bigg[\dfrac{1}{\pi}\int\limits_{-\infty}^{\infty}
\dfrac{x\ln(1+e^{\alpha-x^2})dx}{\tau(x-\tau) \ln(1+e^{\alpha-x^2})}\Bigg].
$$

Следовательно, точное значение скорости скольжения в случае диффузного
рассеяния для квантовых ферми--газов, близким к классическим газам
(т.е. $\alpha\to -\infty$) таково:
$$
U_{sl}(q=1,\alpha=-\infty)=1.0162G_v.
$$

Скорость скольжения в третьем
приближении равна:
$$
U_{sl}(q,\alpha)=G_v \dfrac{2-q}{q}\Big[U_0+U_1q+U_2q^2+U_3q^3\Big],
$$
или, согласно приведенным выше результатам:
$$
U_{sl}(q,\alpha)=G_v
\dfrac{2-q}{q}\Big[0.886227+0.140523q-0.011556q^2+0.001092q^3\Big].
$$

Нетрудно проверить, что в нулевом (максвелловском) приближении
$$
U_0^{sl}(q=1,\alpha=-\infty)=0.886227G_v,
$$
т.е. нулевое приближение дает ошибку
$12.8\%$.

В первом приближении получаем
$$
U_1^{sl}(1)=1.0268G_v,
$$
значит, первое приближение дает ошибку $1.04\%$.

Во втором приближении
$$
U_2^{sl}(q=1,\alpha=-\infty)=1.0152G_v,
$$
т.е. второе приближение дает ошибку $-0.098\%$.

В третьем приближении
$$
U_3^{sl}(q=1,\alpha=-\infty)=1.0163G_v.
$$

Значит, третье приближение приводит к ошибке $0.009\%$.

Приведенное сравнение последовтельных приближений с точным
результатом свидетельствует о высокой эффективности
предлагаемого метода.

\begin{center}
  \item{}\section{Профиль скорости газа в полупространстве и ее значение
  у стенки}
\end{center}

Массовую скорость, отвечающую непрерывному спектру,
разложим по степеням коэффициента диффузности:
$$
U_c(x)=U_c^{(0)}(x)+qU_c^{(1)}(x)+q^2U_c^{(2)}(x)+\cdots.
\eqno{(8.1)}
$$

Тогда профиль массовой скорости в полупространстве можно строить
по формуле:
$$
U(x)=U_{sl}(q,\alpha)+G_vx+U_c(x),
\eqno{(8.2)}
$$
где $U_c(x)$ определяется предыдущим равенством (8.1).

Коэффициенты ряда (8.1) вычислим согласно выведенным выше
формулам:
$$
U_c^{(n)}(x)=G_v\dfrac{2-q}{2\pi}\int\limits_{-\infty}^{\infty}
e^{ikx}E_n(k)dk, \qquad n=0,1,2,\cdots .
$$

Согласно (8.2) вычислим скорость газа непосредственно у стенки:
$$
U(0)=U_{sl}(q,\alpha)+U_c^{(0)}(0)+qU_c^{(1)}(0)+q^2U_c^{(2)}(0)+\cdots.
\eqno{(8.3)}
$$

В случае чисто диффузного отражения молекул от стенки ($q=1$)
согласно (8.3) мы имеем
$$
U(0)=U_{sl}(1,\alpha)+U_c^{(0)}(0)+U_c^{(1)}(0)+U_c^{(2)}(0)+\cdots.
%\eqno{(4.23)}
$$

Отсюда в нулевом приближении
получаем:
$$
U^{(0)}=U_{sl}(1,\alpha)+U_c^{(0)}(0).%=0.674744G_v,
$$

Отсюда видно, что
$$
U^{(0)}\Big|_{\alpha=-\infty}=
U_{sl}(1,-\infty)+U_c^{(0)}(0)\Big|_{\alpha=-\infty}=0.6747G_v.
$$

В первом приближении получаем:
$$
U^{(1)}(0)=U_{sl}(1,\alpha)+U_c^{(0)}(0)+U_c^{(1)}(0).%=0.710319G_v,
$$

Отсюда видно, что
$$
U^{(1)}(0)\Big|_{\alpha=-\infty}=U_{sl}(1,-\infty)+
U_c^{(0)}(0)\Big|_{\alpha=-\infty}+$$$$+U_c^{(1)}(0)\Big|_{\alpha=-\infty}
=0.7103G_v.
$$

Во втором приближении получаем:
$$
U^{(2)}(0)=U_{sl}(1,\alpha)+U_c^{(0)}(0)+U_c^{(1)}(0)+U_c^{(2)}(0).%=0.706802G_v.
$$

Отсюда видно, что
$$
U^{(2)}(0)\Big|_{\alpha=-\infty}=U_{sl}(1, -\infty)+
U_c^{(0)}(0)\Big|_{\alpha=-\infty}+$$$$+U_c^{(1)}(0)\Big|_{\alpha=-\infty}+
U_c^{(2)}(0)\Big|_{\alpha=-\infty}=0.7068G_v.
$$

Сравним эти результаты с точным значение скорости у стенки
\cite{Lat2001TMF}:
$$
U(0,\alpha)=\sqrt{\dfrac{l_2^F(\alpha)}{l_0^F(\alpha)}}G_v.
$$

Из этой формулы вытекает, что
$$
U(0)\Big|_{\alpha=-\infty}=\dfrac{1}{\sqrt{2}}G_v=0.7071G_v.
$$
Введем относительную ошибку
$$
O_n=\dfrac{U(0)-U^{(n)}(0)}{U(0)}\cdot 100\% , \qquad
n=0,1,2,\cdots.
$$

В нулевом приближении относительная ошибка равна $4.6\%$,
в первом приближении равна
$-0.45\%$, во втором приближении равна: $0.044\%$.

%Профили распределения массовой скорости в полупространстве
%построим (см. рис. 1--3), используя формулу (6.21).

\begin{center}\bf
\item{}\section{Приведение формул к размерному виду}
\end{center}

Формулу (6.3) для безразмерной скорости скольжения приведем к
размерному виду. Для этого понадобится коэффициент вязкости
квантового ферми -- газа.

По определению коэффициент кинематической вязкости равен:

$$
\eta=-\dfrac{P_{xy}}{\Big(\dfrac{du_y}{dx}\Big)_\infty},
$$
где $u_y(x)$ -- размерная массовая скорость, откуда

$$
\eta=-\dfrac{m}{g_v}\int fv_xv_y\,d\Omega,
\eqno{(9.1)}
$$
где $g_v$ -- размерный градиент массовой скорости. Учитывая, что
$x_1=\nu \sqrt{\beta}x$, где $x$ -- размерная координата, имеем:
$$
g_v=\Big(\dfrac{du_y(x)}{dx}\Big)_\infty=
\dfrac{\nu d(\sqrt{\beta}u_y(x))}{d(\nu \sqrt{\beta}x)}=
\nu\dfrac{dU_y(x_1)}{dx_1}=\nu G_v.
$$
Здесь $G_v$ -- безразмерный градиент, $U_y$ -- безразмерная массовая
скорость в направлении оси $y$.

Перейдем в (9.1) к интегрированию по безразмерным компонентам
скорости:
$$
\eta=-\dfrac{m^4(2s+1)}{\nu G_v(2\pi\hbar)^3(\sqrt{\beta})^5}
\int fC_xC_y\,d^3C=$$

$$=
-\dfrac{m^4(2s+1)}{\nu G_v(2\pi\hbar)^3(\sqrt{\beta})^5}
\int C_x\,C_y^2\,g_F(C)\,h(x,C_x)\,d^3C.
$$

Подставляя вместо $h(x,C_x)$ асимптотическую функцию
$h_{as}(x,C_x)$, находим, что

$$
\eta=\dfrac{2(2s+1)m^4}{\nu (2\pi\hbar)^3(\sqrt{\beta})^5}
\int C_x^2\,C_y^2\,g_F(C)\,d^3C.
$$

Вычисляя интеграл в этом выражении, находим коэффициент вязкости:
$$
\eta=\dfrac{2(2s+1)m^4\pi l_2^F(\alpha)}{\nu (2\pi\hbar)^3
(\sqrt{\beta})^5},
$$
где
$$ l_2^F(\alpha)=\int\limits_{0}^{\infty}
x^2\ln(1+\exp(\alpha-x^2))\,dx.
$$

Выразим коэффициент вязкости через числовую плотность. Нетрудно
видеть, что
$$
N=\int f\,d\Omega=\dfrac{2\pi (2s+1)m^3l_0^F(\alpha)}
{(2\pi\hbar)^3 (\sqrt{\beta})^3},
$$
$$
l_0^F(\alpha)=\int\limits_{0}^{\infty}\ln(1+\exp(\alpha-x^2))\,dx.
$$

Следовательно, коэффициент вязкости можно представить в виде:
$$
\eta=\dfrac{N\,m\,l_2(\alpha)}{\nu\,\beta\,l_0^F(\alpha)}=
\dfrac{\rho}{\nu\,\beta}\cdot\dfrac{l_2^F(\alpha)}{l_0^F(\alpha)}.
\eqno{(9.3)}
$$

Выражение для размерной скорости с учетом равенства (6.3), в котором
все коэффициенты ряда найдены, перепишем в виде:
$$
\sqrt{\beta}u_{sl}(\alpha,q)=C(\alpha,q)\dfrac{g_v}{\nu},
$$
откуда размерная скорость скольжения равна:
$$
u_{sl}(\alpha,q)=\dfrac{C(\alpha,q)}{\nu \,\sqrt{\beta}\,l}\,l\,g_v.
\eqno{(9.4)}
$$

Здесь
$$
C(q,\alpha)=\dfrac{2-q}{q}\Big[U_0+U_1q+U_2q^2\cdots\Big].
\eqno{(9.5)}
$$

Длину свободного пробега $l$ в (9.4) выразим через вязкость $\eta$ согласно
Черчиньяни \cite{Cerc62}--\cite{8}: $l=\eta \rho^{-1}\sqrt{\pi\beta}$. Подставляя
выражение (9.5) в (9.4), получаем искомую размерную скорость скольжения:
$$
u_{sl}(\alpha,q)=K_v^F(\alpha,q)\,l\,g_v,
%\eqno{(4.4)}
$$
где
$$
K_v^F(\alpha,q)=\dfrac{C(\alpha,q)\,l_0^F(\alpha)}
{\sqrt{\pi}\,l_2^F(\alpha)}
$$
есть коэффициент изотермического скольжения.

\begin{figure}[h]\center
\includegraphics[width=18.0cm, height=12cm]{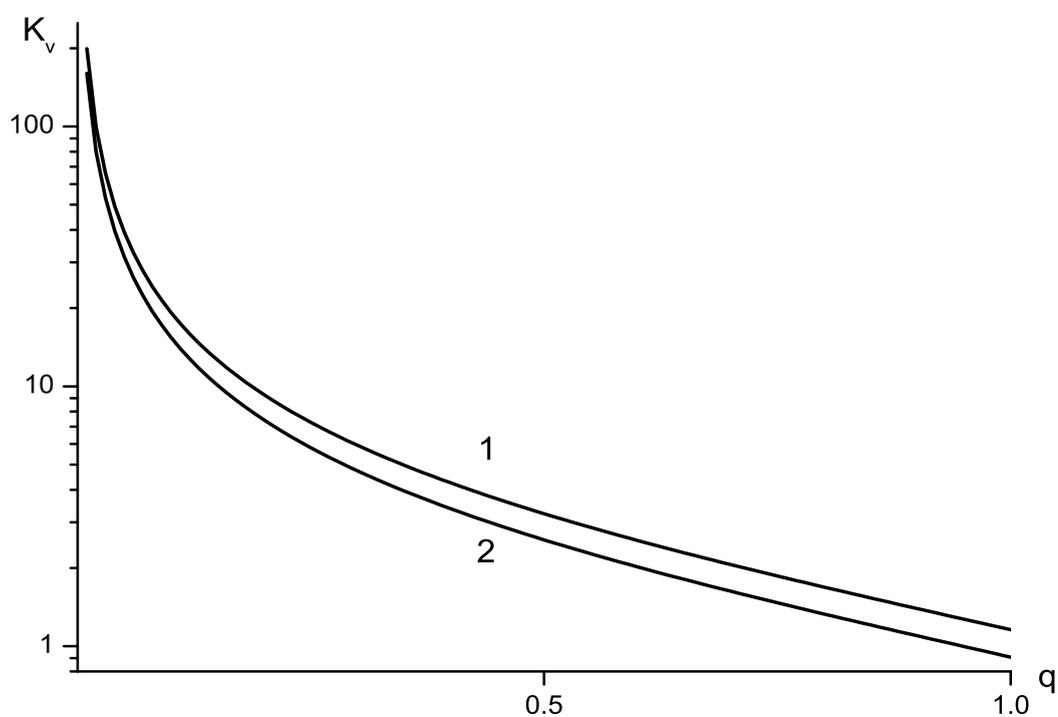}%
\noindent\caption{Зависимость коэффициента изотермического скольжения от
коэффициента диффузности. Кривые $1$ и $2$ отвечают значениям приведенного
химического потенциала $\alpha=-5$ и $\alpha=2$. Логарифмический масштаб по 
вертикальной оси.
}\label{rateIII}
\end{figure}

\begin{figure}[h]\center
\includegraphics[width=18.0cm, height=12cm]{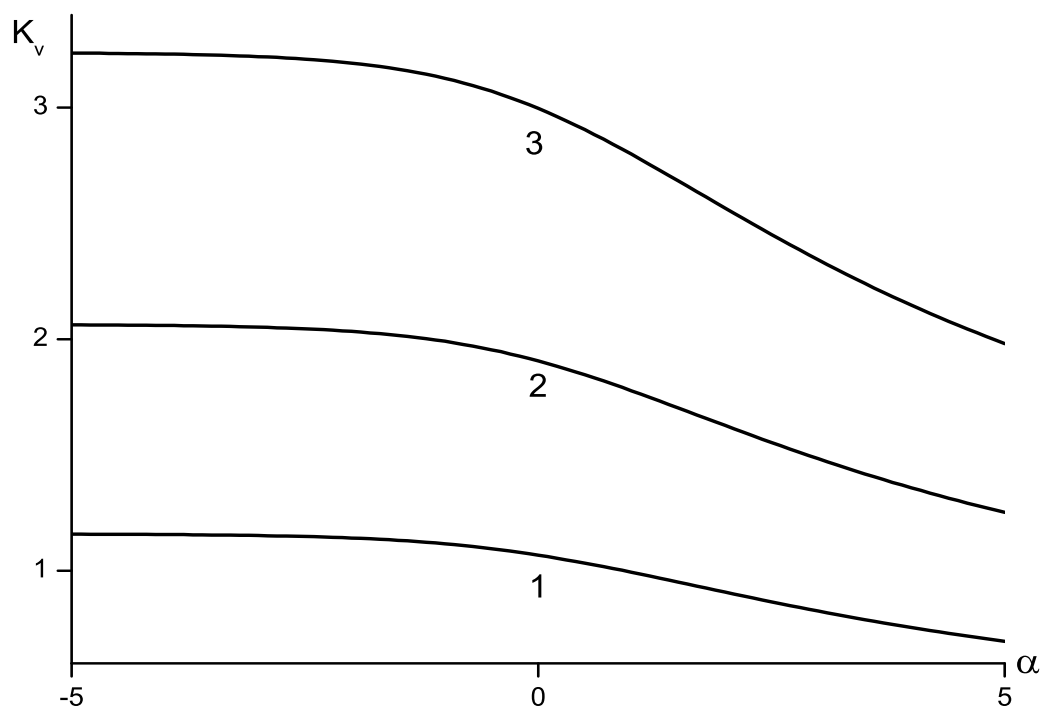}%
\noindent\caption{Зависимость коэффициента изотермического скольжения от
приведенного химического потенциала. Кривые $1,2$ и $2$ отвечают значениям 
коэффициента диффузности $q=1,0.7$ и $q=0.5$. 
}\label{rateIII}
\end{figure}

\begin{center}
  \item{}\section{Заключение}
\end{center}
В настоящей работе с помощью развитого недавно \cite{LatYushk2012}
нового метода решена
полупространственная граничная задача кинетической теории --- задача
Крамерса об изотермическом скольжении квантового ферми--газа с постоянной
частотой столкновений молекул и
с зеркально -- диффузными граничными условиями.
В основе метода лежит идея продолжить функцию распределения в сопряженное
полупространство $x<0$ и включить в кинетическое уравнение
граничное условие в виде члена типа источника
на функцию распределения, отвечающую непрерывному спектру.
С помощью преобразования Фурье кинетическое уравнение сводим к
характеристическому интегральному уравнению Фредгольма
второго рода, которое решаем
методом последовательных приближений.
Для этого разлагаем в ряды по степеням коэффициента диффузности
скорость скольжения газа, его функцию распределения и массовую
скорость, отвечающие непрерывному спектру. Подставляя эти
разложения в характеристическое уравнение и приравнивая
коэффициенты при одинаковых степенях коэффициента диффузности,
получаем счетную систему зацепленных уравнений, из которых
находим все коэффициенты искомых разложений.

Мы находим так называемую скорость скольжения газа вдоль поверхности,
функцию распределения и распределение массовой скорости в
полупространстве. Скорость скольжения --- это фиктивная скорость
газа, которая получается, если профиль асимптотического
распределения массовой скорости, вычисленную вдали от стенки на
основе асимптотического распределения Чепмена --- Энскога,
пролонгировать до границы полупространства.

Предлагаемый метод
обладает высокой эффективностью. Так, сравнение с точным
решением показывает, что в третьем приближении ошибка не
превосходит $0.1\%$.

Кроме того, в работе впервые рассматривается обратная задача
Крамерса. В обратной задаче Крамерса скорость скольжения
считается заданной, а неизвестной величиной является величина
градиента массовой скорости газа как функция коэффициента
диффузности.

Изложенный в работе метод был успешно применен
\cite{53}--\cite{64}
в решении ряда таких сложных задач кинетической теории, которые не
допускают аналитического решения.

%\clearpage
\renewcommand{\baselinestretch}{1.}

\end{document}